\documentclass[aps,prd,showpacs,onecolumn,notitlepage,superscriptaddress,nofootinbib,amsmath,amsfont,amssymb,graphicx,10pt,singlespacing] {revtex4}
\pdfoutput=1
\usepackage{color,graphicx,epsfig}
\usepackage{amsmath} 
\usepackage{amssymb} 
\usepackage{bm}
\usepackage{braket}
\usepackage{slashed}
\usepackage{hyperref}
\usepackage{ifpdf}
\usepackage{latexsym}
\usepackage{bm}

\usepackage[english]{babel}

\definecolor{nicered}{rgb}{0.7,0.1,0.1}
\definecolor{nicegreen}{rgb}{0.1,0.5,0.1}
\hypersetup{colorlinks,citecolor= nicegreen,linkcolor= nicered}

\newcommand {\E}[1]{\times 10^{#1}} 
\newcommand {\e}[1]{\mathrm{~#1}} 
\newcommand{\mc}[1]{\mathcal{#1}}

\newcommand{\mrm}[1]{\mathrm{#1}}

\renewcommand{\Re}[0]{\mrm{Re}}
\renewcommand{\Im}[0]{\mrm{Im}}

\definecolor{Red}{rgb}{1.,0.,0.}

\definecolor{Blue}{rgb}{0.,0.,1.}

\begin{document}

\title{Limits on scalar leptoquark interactions and consequences for
GUTs}

\author{Ilja Dor\v sner} \email[Electronic address:]{ilja.dorsner@ijs.si}
\affiliation{Department of Physics, University of Sarajevo, Zmaja od Bosne 33-35, 71000
  Sarajevo, Bosnia and Herzegovina}

\author{Jure Drobnak} 
\email[Electronic address:]{jure.drobnak@ijs.si} 
\affiliation{J. Stefan Institute, Jamova 39, P. O. Box 3000, 1001 Ljubljana, Slovenia}

\author{Svjetlana Fajfer} \email[Electronic
address:]{svjetlana.fajfer@ijs.si} 
\affiliation{Department of Physics,
  University of Ljubljana, Jadranska 19, 1000 Ljubljana, Slovenia}
\affiliation{J. Stefan Institute, Jamova 39, P. O. Box 3000, 1001
  Ljubljana, Slovenia}

\author{Jernej F. Kamenik} 
\email[Electronic address:]{jernej.kamenik@ijs.si} 
\affiliation{J. Stefan Institute, Jamova 39, P. O. Box 3000, 1001 Ljubljana, Slovenia}
\affiliation{Department of Physics,
  University of Ljubljana, Jadranska 19, 1000 Ljubljana, Slovenia}

\author{Nejc Ko\v snik} 
\email[Electronic address:]{kosnik@lal.in2p3.fr}
\affiliation{Laboratoire de l'Acc\'el\'erateur Lin\'eaire,
Centre d'Orsay, Universit\'e de Paris-Sud XI,
B.P. 34, B\^atiment 200,
91898 Orsay cedex, France}
\affiliation{J. Stefan Institute, Jamova 39, P. O. Box 3000, 1001 Ljubljana, Slovenia}

\date{\today}

\begin{abstract}
  A colored weak singlet scalar state with hypercharge $4/3$ is one of
  the possible candidates for the explanation of the unexpectedly
  large forward-backward asymmetry in $t\bar t$ production as measured
  by the CDF and D\O \, experiments. We investigate the role of this state
  in a plethora of flavor changing neutral current processes and
  precision observables of down-quarks and charged leptons. Our
  analysis includes tree- and loop-level mediated observables in the
  $K$ and $B$ systems, the charged lepton sector, as well as the $Z\to
  b \bar b$ decay width. We perform a global fit of the relevant
  scalar couplings.  This approach can explain the $(g-2)_{\mu}$
  anomaly while tensions among the CP violating observables in the
  quark sector, most notably the nonstandard CP phase (and width
  difference) in the $B_s$ system cannot be fully relaxed.  The
  results are interpreted in a class of grand unified models which
  allow for a light colored scalar with a mass below $1$\,TeV.  We
  find that the renormalizable $SU(5)$ scenario is not compatible with
  our global fit, while in the $SO(10)$ case the viability requires
  the presence of both the $126$- and $120$-dimensional
  representations.
  \end{abstract}


\maketitle

%
\section{Introduction}
\label{sec:intro}
%

Recent CDF and D\O \, results on the forward-backward asymmetry (FBA) in
top quark pair production have attracted a lot of attention and a
number of proposals have been made in order to explain all the
relevant observables (for a recent review see~\cite{topReview}).
Among these, a colored weak singlet scalar with charge $4/3$
$(\overline{\bm{3}}, \bm{1}, 4/3)$, if exchanged in the $u$-channel,
can well accommodate most of the present
measurements~\cite{Dorsner:2009mq,Gresham:2011pa} (see however
also~\cite{Blum:2011fa,Gresham:2011fx}).  Motivated by the success of
this proposal~\cite{Dorsner:2009mq} we have systematically
investigated the role of such state in charm and top quark
physics~\cite{Dorsner:2010cu}. Constraints on the relevant couplings
to up-type quarks come from observables related to $D^0-\bar D^0$
oscillations, as well as di-jet and single top production measurements
at the Tevatron and the LHC. In turn, we were able to predict the
expected rates of flavor changing neutral current~(FCNC) mediated top
quark and charmed meson decays, generated by the presence of the new
colored scalar. Due to the possibility to accommodate such states
within $SU(5)$ grand unified theories (GUTs) that contain $5$- and
$45$-dimensional Higgs representations, we have determined the
resulting textures of the up-quark mass matrix at the GUT scale. The
particular $SU(5)$ model we advocated has an appealing feature of
correlating the presence of light colored scalars stemming from the
45-dimensional Higgs representation with bounds on the partial proton
lifetimes. Namely, the aforementioned representation contains among
other states two colored scalars---$(\overline{\bm{3}}, \bm{1}, 4/3)$
and $(\bm{8}, \bm{2}, 1/2)$---whose masses are below or of the order
of $1$\,TeV when partial proton decay lifetimes are predicted to be at
or slightly above the current experimental bounds. The most common
renormalizable models based on $SO(10)$
framework~\cite{Georgi:1975qb,Fritzsch:1974nn}, on the other hand,
usually rely on inclusion of $120$- and $126$-dimensional scalar
representations to generate fermion masses. As it turns out, both of
these contain a colored weak singlet $(\overline{\bm{3}}, \bm{1},
4/3)$ state that, if light, could accommodate the $t \bar t$
production observables~\cite{Dorsner:2009mq,Patel:2011eh}.  Our
analysis of its couplings could thus help in establishing the
viability of such $SO(10)$ scenarios as we demonstrate later.

During the last decades, rare processes involving down-type quarks and
charged leptons have played an important role in revealing possible
signs of new physics (NP) at low energies. A prominent example is the
anomalous magnetic moment of the muon, whose most precise experimental
measurement~\cite{Bennett:2004pv} deviates from theoretical
predictions within the SM~\cite{Jegerlehner:2007xe} by about three
standard deviations. Also most recently, the CDF and D\O \,
experiments~\cite{Aaltonen:2007he,:2008fj,D0betasUpdate,CDFbetasUpdate}
have reported indications of a large CP-violating phase entering the
$B_s-\bar B_s$ mixing amplitude, which cannot be accommodated within
the standard Cabibbo-Kobayashi-Maskawa~(CKM) framework.\footnote{The
  recent D\O \, measurement of an anomalous like-sign dimuon charge
  asymmetry~\cite{Abazov:2010hv,Abazov:2011yk} is consistent with the
  hypothesis of NP contributing only in $B_{d,s}-\bar B_{d,s}$
  mixing~\cite{Ligeti:2010ia,Lenz:2010gu}.}  {Furthermore, the
  experimental data seem to prefer the decay widths' difference
  between the $B_s$ eigenstates $\Delta \Gamma_s$ } larger than
  predicted within the SM. {Any NP addressing this discrepancy would
  have to contribute to the absorptive part of the mixing
  amplitude~\cite{Grossman:1996era}.
  Finally, the recent measurements of the leptonic $B\to \tau\nu$
  branching ratio induce a $2.9\,\sigma$ tension on the global fit to
  the CKM unitarity triangle which may be ameliorated via a small NP
  contribution to the $B_d-\bar B_d$ mixing phase~\cite{Lenz:2010gu}.
}

Motivated by the interesting role the scalar $(\overline{\bm{3}},
\bm{1}, 4/3)$ state might play in down-quark and charged lepton
physics, we systematically investigate its contributions to the
down-quark and charged lepton physical observables.  We consider
observables affected already at the tree-level, as for example
$K_{L,S} (B_{s,d}) \to \ell^+ \ell^{'-}$, lepton flavor violating
(LFV) tau decays and $\mu-e$ conversion in nuclei, and also loop
suppressed contributions to $\epsilon_K$, $\Delta m_K$,
$B_{d,s}-\bar B_{d,s}$ oscillation observables, lepton anomalous
magnetic moments, LFV radiative tau and muon decays, and the $Z \to b
\bar b$ decay width. 
These constraints can be consistently implemented
within the framework of $SU(5)$ and $SO(10)$ GUTs which rely on
tree-level generation of charged fermion masses. 

The outline of this article is as follows. In Sec.~\ref{sec:ewframe}
we define the leptoquark couplings of the scalar $(\bar{\bm
  3},\bm{1},4/3)$ to SM fermions. In Sec.~\ref{sec:pheno} we study
the effects of this state on low energy precision observables and perform a
global fit of its couplings in Sec.~\ref{sec:fit}. Resulting values of couplings are then
reanalyzed in Sec.~\ref{sec:gutframework},
where we study how they relate to the mechanism of mass generation in
GUT scenarios and derive bounds on the vacuum expectation value of the
45-dimensional Higgs representation in the $SU(5)$ case. Finally we conclude in
Sec.~\ref{sec:conclusions}.

%
\section{Electroweak scale framework}
\label{sec:ewframe}
%

We consider a color triplet, weak singlet scalar with charge $4/3$
\begin{equation}
\Delta = (\overline{\bm{3}}, \bm{1}, 4/3)\,,
\end{equation}
which can couple to the right-handed fermions of the SM via the interaction Lagrangian
\begin{equation}
\label{eq:lagr}
  \mc{L}_{\Delta} =  Y_{ij} \bar{\ell}_{i} P_L
  d^C_{ja} \Delta^{a*}+ \frac{g_{ij}}{2} \epsilon_{abc}
  \bar{u}_{ia} P_L u^C_{jb} \Delta^c + \rm{h.c.}\,,
\end{equation}
where $P_{L,R} = (1\mp \gamma_5)$ and the totally antisymmetric tensor
$\epsilon_{abc}$ is defined with $\epsilon_{123}=1$. The above
interaction terms exhaust all possibilities of renormalizable
couplings between SM fermions and the $\Delta$ scalar. As mentioned in
Sec.~\ref{sec:intro}, the diquark couplings $g_{ij}$ of $\Delta$ to
up-quarks of different generations $u_i u_j$ can play an important
role in top and charm physics. The leptoquark nature of $\Delta$, on
the other hand, is parameterized by couplings $Y_{ij}$ to charged
leptons and down-quarks, $\ell_i d_j$.  If, and only if both $g$ and
$Y$ are present, baryon~($B$) and lepton~($L$) numbers are violated
while their combination $B-L$ is conserved. Proton lifetime is
protected in this general case by antisymmetric color contraction
between $\Delta$ and two up-quarks. It implies flavor-space
antisymmetric coupling $g_{ij}(=-g_{ji})$ which prohibits the proton
from decaying via dimension-6 effective operators mediated by
$\Delta$. To comply with the $t \bar t$ production parameters, the
mass of $\Delta$ should be below $1$\,TeV, preferably around
$400$\,GeV.  This setup is natural in a theoretically well-motivated
class of grand unified models. A realistic GUT context
of~\eqref{eq:lagr} and the resulting model building constraints will
be presented in Sec.~\ref{sec:gutframework}.

In general $Y$ is a complex matrix acting on charged-lepton
and down-quark flavor indices
$
  \bar \ell_R Y (d_R)^C\,.
$
From the right-hand side of $Y$ one may redefine the quark fields
using the global $B$ symmetry transformation, which since it is
broken, has a side effect of globally rephasing diquark couplings $g$
while leaving the mass and CKM matrices invariant. This is not a worry
at this point since current experimental constraints from top quark
physics and $D^0-\bar D^0$ mixing observables cannot probe the
overall phase of $g$~\cite{Dorsner:2010cu}. One can redefine lepton
fields in an analogous manner. However, of the two independent phases
used to redefine quark and lepton fields only their sum is physical,
while their difference corresponds to $B-L$, a conserved quantum
number. As a result, freedom remains to choose one phase in
$Y$. Strictly speaking, from the phenomenological point of view in
Sec.~\ref{sec:pheno} where we do not consider observables sensitive
to lepton mixing, we could have rephased charged lepton flavors
independently. This would allow us to rephase each row of $Y$
separately. We restrain however from using this freedom which would result
in the leptonic phase convention being ``gauged'' according to $Y$ instead
of to the standard form of the Pontecorvo-Maki-Nakagawa-Sakata~(PMNS) matrix.

%
\section{Leptoquark probing observables}
\label{sec:pheno}
%

The leptoquark couplings endow the scalar $\Delta$ with a potential to
cause large effects in (flavor changing) neutral current processes of
down quarks and charged leptons (see~\cite{Saha:2010vw} for a recent analysis of scalar leptoquark constraints from $K$ and $B$ sectors). The couplings $Y_{ij}$ of
Eq.~\eqref{eq:lagr} must therefore pass constraints from plethora of
precisely measured or bounded low energy observables. In this section
we make predictions of the observables most sensitive to effects
stemming from $\Delta$ and compare them to current experimental
values. For each observable we state an effective error of the
constraint, which is, as will be explained in the following, a total
combined theoretical and experimental uncertainty. In order to
confront this model with experimental data in a quantitative manner we
wrap up the analysis with a global fit of all the $9$ entries of $Y$
in Sec.~\ref{sec:fit}.

\subsection{SM theoretical inputs}
\label{sec:CKM}

Most observed flavor phenomena are well described within the SM and
thus the allowed size of NP contributions crucially depends on
reliable estimates of SM parameters. In the presence of NP virtual
contributions to quark flavor observables the extraction of the CKM
matrix becomes more involved, since some observables used in the
conventional CKM fits receive contributions from both the SM and NP
amplitudes. As we want to treat the SM contributions as a theoretical
background, it is imperative to calibrate the CKM matrix exclusively
on SM tree-level observables, which are largely insensitive to virtual
$\Delta$ contributions.

Thus we employ the results of a simple CKM fit to tree-level
observables.\footnote{We do not use available results in the
  literature since they do not provide correlations among the
  parameters~(e.g.\,UTFit tree-level fit~\cite{UTfit}).} These are the
measurements of the first and the second row CKM element moduli from
super-allowed $\beta$ decays, leptonic and semileptonic meson decays,
as well as
 the extraction of the CP phase angle $\gamma$ from tree-dominated $B$
 decays~\cite{Nakamura:2010zzi}
\begin{equation}
  \label{eq:ckminputs}
  |V_\mrm{CKM}| = \begin{pmatrix}
    0.97425(22) & 0.2252(9) & 3.89(44)\E{-3}\\
    0.23(11)\phantom{000} & 1.023(36) & 4.06(13)\E{-2}\\
    & &  \phantom{0.88(7) 1\E{-2}}
  \end{pmatrix}\,,
\qquad
\gamma = 73({}^{+22}_{-25})\,{}^\circ\,.
\end{equation}
{In particular, the value of $|V_{ub}|$ is an
   average of exclusive and inclusive semileptonic $B$ decay analyses. We explore the impact of the branching ratio of $B \to
   \tau \nu$ on the CKM fit in Sec.~\ref{sec:Vub}.}
Note that we cannot use direct $|V_{tb}|$ determination from single
top production measurements, since these may be affected by $\Delta$
contributions~\cite{Dorsner:2010cu}. By fitting
constraints~\eqref{eq:ckminputs} to the Wolfenstein expanded CKM
matrix up to order $\lambda^4$, we find values in agreement
with~\cite{UTfit}
\begin{align}
\label{eq:CKMcv}
\lambda &= 0.22538(65)\,,\\
A &=0.799(26)\,,\nonumber\\
\rho &=0.124(70)\,,\nonumber\\
\eta &=0.407(52)\,,\nonumber
\end{align}
while we also extract the correlation matrix between the fit parameters
\begin{equation}
  \label{eq:corr}
  \begin{pmatrix}
  1 &&&\\
 -0.178 & 1 &  & \\
 -0.00517 & -0.0553 & 1 & \\
 -0.0226 & -0.242 & -0.198 & 1
\end{pmatrix}\,.
\end{equation}
In addition, we use the top quark pole mass of $m_t =
173.3$\,GeV~\cite{:1900yx}, and the $\overline{\rm MS}$ bottom and
charm quark masses $m_b(m_b) = 4.2$\,GeV, and $m_c(m_c) =
1.29$\,GeV~\cite{Nakamura:2010zzi}.
Observable-specific numerical inputs will be stated where needed.

\subsection{Tree-level constraints}
\label{sec:treelevel}

We first focus our attention on observables which receive possible $\Delta$ contributions already at the tree-level and thus represent potentially most severe constraints on the $Y$ matrix.
The relevant effective Lagrangian for processes involving charged lepton and
down-quark pairs results from integrating out the $\Delta$ at tree level. After applying
Fierz identities we recover the LFV and quark FCNC interaction terms
among the right-handed leptons and quarks
\begin{equation}
  \label{eq:Heff}
  \mc L^{\Delta}_{d_i \bar d_j \to \ell_a^- \ell_b^+} =
  -\frac{Y_{aj} Y_{bi}^*}{2 m_\Delta^2} (\bar \ell_a \gamma^\mu P_R \ell_b)( \bar d_j \gamma_\mu P_R d_i)\,.
\end{equation}
The corresponding leptonic (LFV) decay width of a neutral
pseudoscalar meson $P (d_i \bar d_j) \to \ell_a^- \ell_b^+$ is given by
\begin{equation}
  \label{eq:leptonicwidth}
  \Gamma_{P(d_i \bar d_j) \to \ell^-_a \ell^+_b} = \frac{|Y_{aj} Y_{bi}^*|^2}{512\pi} \frac{m_P^3 f_P^2}{m_\Delta^4} \left[\hat
    m_a^2 + \hat m_b^2-(\hat m_a^2 - \hat m_b^2)^2 \right]
  \left[(1-(\hat m_a+\hat m_b)^2)(1-(\hat m_a-\hat m_b)^2)\right]^{1/2},
\end{equation}
where $m_P$ is the decaying meson mass, its decay constant is defined
as customary for light neutral mesons ($\pi^0, K^0$), $\Braket{0 | \bar d_j
  \gamma^\mu \gamma_5 d_i | P(p)} = i p^\mu f_P/\sqrt{2}$, while the
hatted masses of leptons are $\hat m_{a,b} = m_{a,b}/m_P$. We study the particularly interesting decay modes below.

\subsubsection{$K_L \to \mu^- \mu^+,\, e^+ e^-,\, \mu^\pm e^\mp$}

While the decay $K_L \to \mu^- \mu^+$ has been measured with great precision
($\mc{B} = (6.84\pm0.11)\E{-9}$~\cite{Nakamura:2010zzi}), the
presence of long-distance intermediate states $K_L \to \gamma^*
\gamma^* \to \mu^+ \mu^-$ precludes similarly reliable SM predictions for this observable. We use a
conservative estimate for the pure short distance branching fraction
$\mc{B}^\mrm{exp}_\mrm{SD} < 2.5\E{-9}$, obtained using dispersive
techniques~\cite{Isidori:2003ts,D'Ambrosio:1997jp}, as a $1\,\sigma$
upper bound. Since the SM short distance contribution
$\mc{B}_\mrm{SM(SD)} \approx 0.9\E{-9}$ is much smaller, we can
neglect it
and keep only the $\Delta$-mediated amplitude. For the decay width $K_L
\to \mu^+ \mu^-$, CP violation in $K-\bar K$ mixing is irrelevant
and we treat $K_L$ as a pure CP-odd state. Contributions of both $K^0$
and $\bar K^0$ amplitudes are to be taken into account using
Eq.~\eqref{eq:leptonicwidth} by replacing $Y_{aj} {Y_{bi}}^* \to
\sqrt{2} \Re (Y_{\mu s} {Y_{\mu d}}^*)$. The decay width, mediated by
CP conserving combination of couplings $Y$, is then
\begin{equation}
  \label{eq:KLmumu}
 \Gamma_{K_L \to \mu^- \mu^+} =\frac{\left[\Re
     \left(Y_{\mu s} Y_{\mu d}^*\right)\right]^2}{128\pi} \frac{m_K^3
   f_K^2}{m_\Delta^4} \hat m_\mu^2 \sqrt{1-4 \hat m_\mu^2}\,.
\end{equation}
Using the lattice value of the kaon decay constant $f_K =
156.0$\,MeV~\cite{Laiho:2009eu}, the numerical result for the
1$\,\sigma$ upper bound is
\begin{equation}
  \label{eq:KLmumulimit}
\left[\Re \left(Y_{\mu s} Y_{\mu d}^*\right)\right]^2 < 2.7\E{-9} \left(\frac{m_\Delta}{400\e{GeV}}\right)^4\,.
\end{equation}

In the di-electron mode  $K_L\to e^+ e^-$ the experimental measurement of $\mathcal B = (9^{+6}_{-4})\times 10^{-12}$~\cite{Nakamura:2010zzi} agrees well with the long-distance dominated SM estimate of $\mathcal B_{\rm LD}^{\rm SM} = (9\pm 0.5)\times 10^{-12}$~\cite{Valencia:1997xe}. The $\Delta$ contribution to this decay mode can be obtained from~\eqref{eq:KLmumu} by replacing $\mu$ with $e$ everywhere. Saturating the experimental uncertainty leads to the following $1\,\sigma$ constraint
\begin{equation}
  \label{eq:KLeelimit}
\left[\Re \left(Y_{e s} Y_{e d}^*\right)\right]^2 < 2.5\E{-7} \left(\frac{m_\Delta}{400\e{GeV}}\right)^4\,.
\end{equation}

A much stronger upper bound of the LFV decays $\mc{B}(K_L \to \mu^\pm
e^\mp)<4.7\E{-12}$ at $90\%\,$ confidence level (C.L.) has been
set in~\cite{Nakamura:2010zzi}. The corresponding form of
Eq.~\eqref{eq:leptonicwidth} is obtained by adding first coherently
the flavor components of $K_L$ and then summing over the widths of the two
oppositely-charged final states. The result, with $m_e$ set to zero,
is
\begin{equation}
  \label{eq:KLmue}
   \Gamma_{K_L \to \mu^\pm e^\mp} =\frac{
     \left|Y_{\mu s} Y_{e d}^* +Y_{\mu d} Y_{e s}^*\right|^2}{512\pi} \frac{m_K^3
   f_K^2}{m_\Delta^4} \hat m_\mu^2 \left[1-\hat m_\mu^2\right]^2\,,
\end{equation}
and implies a $1\,\sigma$ bound
\begin{equation}
  \label{eq:KLmuelimit}
  \left|Y_{\mu s} Y_{e d}^* +Y_{\mu d} Y_{e s}^*\right|^2 <
  1.2\E{-11} \left(\frac{m_\Delta}{400\e{GeV}}\right)^4\,.
\end{equation}

\subsubsection{$K_S \to e^- e^+,\mu^+\mu^-$}

Since $K_S$ is approximately CP-even and is decaying to a CP-odd
final state this decay mode is sensitive to the imaginary parts of
$Y$.  In the muonic channel, the best limit still comes from the early seventies with $\mathcal B(K_S\to \mu^+ \mu^-) < 3.2\E{-7}$ at $90$\,\% C.L.~\cite{Nakamura:2010zzi}, while the best upper bound on the branching
fraction $\mathcal B(K_S\to e^+ e^-) < 9\E{-9}$ at $90$\,\% C.L. was more recently set by the KLOE
experiment~\cite{Ambrosino:2008zi}. Both are still far above the SM
expectations, whose long distance effects through $K_S \to \gamma^*
\gamma^* \to e^- e^+ (\mu^+\mu^-)$ reach $8\E{-9} (2\E{-6}) \times \mc{B} (K_S \to \gamma
\gamma) \sim 10^{-14} (10^{-11})$~\cite{Ecker:1991ru,Nakamura:2010zzi}. These
observables thus present clean probes of CP violating effects in the
effective Lagrangian~\eqref{eq:Heff}, through the decay widths
\begin{eqnarray}
 \label{eq:1}
 \Gamma_{K_S \to e^- e^+} &=&\frac{\left[\Im \left(Y_{e s}
       Y_{e d}^*\right)\right]^2}{128\pi} \frac{m_K^3
   f_K^2}{m_\Delta^4} \hat m_e^2\,,\\
 \Gamma_{K_S \to \mu^- \mu^+} &=&\frac{\left[\Im
     \left(Y_{\mu s} Y_{\mu d}^*\right)\right]^2}{128\pi} \frac{m_K^3
   f_K^2}{m_\Delta^4} \hat m_\mu^2 \sqrt{1-4 \hat m_\mu^2}\,.
\end{eqnarray}
The resulting bounds, although diluted by helicity
suppression in the electron mode and the short lifetime of $K_S$, are important constraints to be
fulfilled by the following combinations of couplings at $1\,\sigma$ C.L.
\begin{eqnarray}
 \left[\Im \left(Y_{e s} Y_{e d}^*\right)\right]^2 &<& 0.13 \left(\frac{m_\Delta}{400\e{GeV}}\right)^4\,, \\
\left[\Im \left(Y_{\mu s} Y_{\mu d}^*\right)\right]^2 &<& 1.1\E{-4} \left(\frac{m_\Delta}{400\e{GeV}}\right)^4\,.
\end{eqnarray}

\subsubsection{$B_{d(s)} \to \ell^- \ell^+$}

In the SM these FCNC processes suffer additional helicity-suppression
($m_\ell^2/m_B^2$) leading to branching fractions of the modes with
electrons, which are negligibly small compared to the current
sensitivities of experiments, as given by the $90$\,\% C.L. upper bounds
on $\mc{B}(B_d \to e^- e^+) < 8.3\E{-8}$ and $\mc{B}(B_s \to e^- e^+)
< 2.8\E{-7}$~\cite{Nakamura:2010zzi}. In the dimuon channel the SM
predictions for the branching fractions---of order $\sim 10^{-10}$
($10^{-9}$) for $B_d$ ($B_s$) decays---are closer to but still an
order of magnitude below current experimental 90\,\% C.L. upper bounds
$4.2\E{-9}$~($1.2\E{-8}$)~\cite{LHCbEPS}. Even in the case of
$B_d \to \tau^- \tau^+$ where the helicity suppression is the least severe, the
SM prediction of $\mc{B} \sim 10^{-7}$~\cite{Harrison:1998yr} is far
below the current experimental reach of $4.1\E{-3}$ at 90\,\%
C.L.~\cite{Aubert:2005qw}.  Consequently we do not need to consider pure
SM or interference terms between SM and $\Delta$-mediated amplitudes
and focus our attention only to the pure $\Delta$ contributions.

\begin{table}[tb]
  \centering
  \begin{tabular}[c]{ccccr@{$\,\,<\,\,$}l}
    \hline
     decay mode && $90$\,\% C.L. exp. bound on $\mc B$ &&
     \multicolumn{2}{c}{$1\,\sigma$ upper bound in units $(m_\Delta/400\e{GeV})^4$} \\\hline
$B_d \to e^- e^+$ && $8.3\E{-8}$&&$\left|Y_{eb} Y_{ed}^*\right|^2$&$4.4$\\
$B_d \to \mu^- \mu^+$ && $4.2\E{-9}$ &&$\left|Y_{\mu b} Y_{\mu d}^*\right|^2$&$5.0\E{-6}$\\
$B_d \to \tau^- \tau^+$ && $4.1\E{-3}$ &&$\left|Y_{\tau b} Y_{\tau d}^*\right|^2$&$1.3\E{-2}$\\
$B_s \to e^- e^+$ && $2.8\E{-7}$ &&$\left|Y_{eb} Y_{es}^*\right|^2$&$10.1$\\
$B_s \to \mu^- \mu^+$ && $1.2\E{-8}$ &&$\left|Y_{\mu b} Y_{\mu
    s}^*\right|^2$&$1.1\E{-5}$\\
    \hline
$B_d \to e^\mp \mu^\pm$ && $6.4\E{-8}$&&$\left|Y_{eb} Y_{\mu
    d}^*\right|^2 + \left|Y_{\mu b} Y_{e d}^*\right|^2$&$1.6\E{-4}$\\
$B_d \to \mu^\mp \tau^\pm$ && $2.2\E{-5}$&&$\left|Y_{\mu b} Y_{\tau
    d}^*\right|^2 + \left|Y_{\tau b} Y_{\mu d}^*\right|^2$&$2.2\E{-4}$\\
$B_d \to \tau^\mp e^\pm$ && $2.8\E{-5}$&&$\left|Y_{\tau b} Y_{e
    d}^*\right|^2 + \left|Y_{e b} Y_{\tau d}^*\right|^2$&$2.7\E{-4}$\\
$B_s \to e^\mp \mu^\pm$ && $2.0\E{-7}$&&$\left|Y_{e b} Y_{\mu
    s}^*\right|^2 + \left|Y_{\mu b} Y_{e s}^*\right|^2$&$3.4\E{-4}$\\
    \hline
  \end{tabular}
  \caption{Limits on $Y$ couplings coming from upper bounds of
    lepton flavor conserving and violating $B_{d(s)} \to \ell^- \ell^+$ decays~\cite{Nakamura:2010zzi,LHCbEPS}.}
  \label{tab:Bll}
\end{table}
In Eq.~\eqref{eq:leptonicwidth} we substitute $f_P \to \sqrt{2}
f_{B_{d(s)}}$ in order to conform with the standard normalization of
heavy pseudoscalar decay constants. The lepton flavor conserving decay
widths then read
\begin{equation}
\Gamma_{B_{d(s)} \to \ell^- \ell^+} =\frac{\left|Y_{\ell
      b} Y_{\ell d(s)}^*\right|^2}{128\pi} \frac{m_{B_{d(s)}}^3
   f_{B_{d(s)}}^2}{m_\Delta^4} \hat m_\ell^2 \sqrt{1-4\hat m_\ell^2}\,,
\end{equation}
while the rates of LFV decays are, e.g., for the $\mu \tau$ final
state
\begin{equation}
\Gamma_{B_{d} \to \tau^- \mu^+} =\frac{\left|Y_{\tau
      b} Y_{\mu d}^*\right|^2}{256\pi} \frac{m_{B_{d}}^3
   f_{B_{d}}^2}{m_\Delta^4} \hat m_\tau^2 (1-\hat m_\tau^2)\,.
\end{equation}
For the other dilepton LFV decays one should adapt the lepton indices of
$Y$ and replace $\hat m_\tau$ with the mass of the heaviest lepton in the
final state.  For the decay constants we use the central
values of recent lattice QCD averages~\cite{Laiho:2009eu}: $f_{B_d} =
193$\,MeV, $f_{B_s} = 239$\,MeV. The compilation of experimental upper
bounds and their resulting interpretation as constraints on $Y$ are
given in Tab.~\ref{tab:Bll}.

\subsubsection{$B\to X_s \ell^+ \ell^-$}
Effective Lagrangian \eqref{eq:Heff} also contributes to the non-helicity suppressed $b\to s \ell^+\ell^-$ transitions. In particular it contributes to the $C_{9'}$ and $C_{10'}$ Wilson coefficients of the effective weak Hamiltonian as defined in~\cite{DescotesGenon:2011yn}. Following this reference, we write
\begin{equation}
\mc{H}^{(bs)}_{\rm eff} = -\frac{4 G_F}{\sqrt 2} \lambda^{(s)}_t \sum_i C^\ell_i \mathcal O^\ell_i\,,
\end{equation}
where $\lambda^{(s)}_t = V_{tb} V_{ts}^*$, and $\Delta$ only contributes to
\begin{equation}
\mathcal O^\ell_{9'} = \frac{e^2}{16\pi^2} (\bar s \gamma_\mu P_R b) (\bar \ell \gamma^\mu \ell)\,, \qquad \mathcal O^\ell_{10'} = \frac{e^2}{16\pi^2} (\bar s \gamma_\mu P_R b) (\bar \ell \gamma^\mu\gamma_5 \ell)\,,
\end{equation}
at the tree-level with the weak-scale Wilson coefficients
\begin{equation}
C^\ell_{9'} = C^\ell_{10'} = - \frac{\sqrt 2 \pi Y_{\ell b} Y_{\ell s}^*}{4 G_F \lambda^{(s)}_t \alpha\, m_\Delta^2 }\,.
\end{equation}
The running of the $O^\ell_{9',10'}$ operators from the weak matching scale to the $b$-quark mass scale is dominated by electroweak effects~\cite{Bobeth:2003at} and can be safely neglected for our purpose.
At present the most sensitive observable is the inclusive decay width of $B\to X_s \ell^+ \ell^-$, where $\ell=e,\mu$, integrated in the dilepton invariant mass range of $m_{\ell^+\ell^-} \equiv \sqrt{(p_{\ell^+}+p_{\ell^-})^2} \in [1,6]$\,GeV\,. The corresponding branching fraction ($\mathcal B_{(1-6)\rm GeV}$) is known in the SM to $10\%$ accuracy and can be written in presence of $C_{9',10'}$ contributions as~\cite{DescotesGenon:2011yn}
\begin{equation}
\mathcal B^{\rm th}_{(1-6)\rm GeV} =  \left| \frac{\lambda_t^{(s)}/V_{cb}}{0.981} \right|^2 [(15.86\pm1.51) - 0.049 {\rm Re}(C^\ell_{9'}) +0.061 {\rm Re}(C^\ell_{10'})  + 0.534 |C^\ell_{9'}|^2 + 0.543 |C^\ell_{10'}|^2] \times 10^{-7}\,.
\end{equation}
The experimental measurements of this quantity by the BaBar~\cite{Aubert:2004it} and Belle~\cite{Iwasaki:2005sy} experiments, averaged over the muon and electron flavors, yield $\mathcal B_{(1-6)\rm GeV}^{\rm exp} = (1.60\pm 0.5)\times 10^{-6}$~\cite{Huber:2007vv}.

\subsubsection{$B\to \pi \ell^+ \ell^{\prime-}$ and $B\to K \ell^+ \ell^{\prime-}$}

The exclusive $B\to \pi \ell^+ \ell^-$ mode, where $\ell=\mu,e$, is severely CKM suppressed in the SM leading to branching ratio predictions which are well below the present experimental bound $\mathcal B(B^+ \to \pi^+ \ell^- \ell^+)<4.9\E{-8}\, @ \,90\%$~C.L.~\cite{Nakamura:2010zzi}. In addition, several LFV $B^+ \to \pi^+(K^+) \ell^+ \ell^{\prime -}$  modes have also been searched for at the $B$-factories and we compile the present bounds in Tab.~\ref{tab:BKll}.
\begin{table}[tb]
  \centering
  \begin{tabular}[c]{ccccr@{$\,\,<\,\,$}l}
    \hline
     decay mode && $90$\,\% C.L. exp. bound on $\mc B$ &&
     \multicolumn{2}{c}{$1\,\sigma$ upper bound in units $(m_\Delta/400\e{GeV})^4$} \\\hline
$B^+ \to \pi^+ \ell^- \ell^+$ && $4.9\E{-8}$&&$\left|Y_{eb} Y_{ed}^*\right|^2+\left|Y_{\mu b} Y_{\mu d}^*\right|^2$&$3.0\E{-7}$\\
$B^+ \to \pi^+ e^\pm \mu^\mp$ && $1.7\E{-7}$&&$\left|Y_{eb} Y_{\mu d}^*\right|^2+\left|Y_{\mu b} Y_{e d}^*\right|^2$&$1.1\E{-6}$\\
$B^+ \to K^+ e^\pm \mu^\mp$ && $9.1\E{-8}$&&$\left|Y_{eb} Y_{\mu s}^*\right|^2+\left|Y_{\mu b} Y_{e s}^*\right|^2$&$4.3\E{-7}$\\
$B^+ \to K^+ \tau^\pm \mu^\mp$ && $7.7\E{-5}$&&$\left|Y_{\tau b} Y_{\mu s}^*\right|^2+\left|Y_{\mu b} Y_{\tau s}^*\right|^2$&$5.7\E{-4}$\\
    \hline
  \end{tabular}
  \caption{Limits on $Y$ couplings coming from upper bounds on $B^+ \to \pi(K) \ell^- \ell^{\prime +}$ branching fractions, compiled by~\cite{Nakamura:2010zzi}.}
  \label{tab:BKll}
\end{table}

The computation of the $\Delta$ contributions to these exclusive rare semileptonic $B$ decays requires the knowledge of the relevant hadronic $\langle \pi | \mathcal J^\mu_{d} | B\rangle$ and $\langle K | \mathcal J^\mu_{s} | B\rangle$ matrix elements, where $\mathcal J_{q}^\mu = \bar b \gamma_\mu P_R q$ is the relevant quark current operator. We employ the form factor parametrization
\begin{eqnarray}
\bra{P_q(p')} \bar b \gamma^\mu q \ket{B(p)} &=& f^{P_q}_+(s) \left[ (p+p')^\mu - \frac{m_B^2-m_{P_q}^2}{s} (p-p')^\mu \right] + f^{P_q}_0(s) \frac{m_B^2-m_{P_q}^2}{s} (p-p')^\mu\,, 
\end{eqnarray}
for $P_d \equiv \pi $ and $P_s \equiv K$, where $s = (p-p')^2$. The $f_{+,0}^{K}$ form factors have been computed using QCD sum rules techniques and we employ the results of~\cite{Ball:2004ye}. For the  $f_{+,0}^{\pi}$ form factors we use a more recent calculation~\cite{Khodjamirian:2011ub}. The $B\to K \tau^{\pm} \mu^{\mp}$ differential decay rate can be written in a compact form by neglecting the small muon mass
\begin{equation}
\frac{d\Gamma}{ds} (B\to K \tau^{\pm} \mu^{\mp}) = \frac{|Y_{\mu s}Y_{\tau b}^*|^2+|Y_{\mu b}Y_{\tau s}^*|^2}{(16 \pi)^3 m_\Delta^4} m_B^3 \lambda^{1/2}  \left(1- \frac{m_\tau^2}{s}\right)^2 \left[\frac{\lambda}{3}  {f_+^K}(s)^2 \left(2+\frac{m_\tau^2}{s}\right)+ \frac{m_\tau^2}{s} {f_0^K}(s)^2 \left(1-\frac{m_K^2}{m_B^2}\right)^2\right]\,,
\end{equation}
where $\lambda \equiv \lambda(1,m_K^2/m_B^2,s/m_B^2)$ and $\lambda(a,b,c) = a^2 + b^2 + c^2 - 2( ab + bc + ca)$\,. For the modes without tau leptons in the final state one can neglect lepton masses completely, i.e.
\begin{equation}
\frac{d\Gamma}{ds} (B\to K e^{\pm} \mu^{\mp}) = \frac{|Y_{\mu s}Y_{e b}^*|^2+|Y_{\mu b}Y_{e s}^*|^2}{ (16 \pi)^3 m_\Delta^4} m_B^3 \lambda^{3/2}   \frac{2}{3}f_+^K(s)^2\,.
\end{equation}
The modes with a pion in the final state can then be simply obtained from the above formula by replacing $s$ with $d$ and $K$ with $\pi$\,. Integrating over the available phase space and comparing to the experimental upper bounds on $B^+ \to \pi(K) \ell^- \ell^{\prime +}$ decays~\cite{Nakamura:2010zzi}, we obtain the constraints listed in Tab.~\ref{tab:BKll}. Finally we note that the corresponding rare $K\to \pi \ell^+ \ell^{\prime -}$ decay modes are always less sensitive to the relevant $Y$ entries compared to the rare leptonic $K_{L,S}\to \ell^+ \ell^{\prime -}$ modes~\cite{Saha:2010vw}.

\subsubsection{LFV semileptonic $\tau$ decays}

These decays constitute important observables, uniquely sensitive to the third
row of $Y$. Upper limits on their branching fractions have been
set by the Belle and BaBar experiments. The width of the pionic channel reads
\begin{equation}
  \label{eq:tauwidth}
  \Gamma_{\tau \to \ell \pi^0} = \frac{\left|Y_{\ell d} Y_{\tau
        d}^*\right|^2}{2048\pi} \frac{f_\pi^2 m_\tau^3}{m_\Delta^4}
  \left[1-3 \hat m_\ell^2 - 2 \hat m_\pi^2 \right]\,,
\end{equation}
where we have kept the leading powers of final state particle
masses. Decay width for a channel with $K_S$ in the final state is
obtained from~\eqref{eq:tauwidth} by replacing $Y_{\ell d} Y_{\tau
  d}^* \to Y_{\ell d} Y_{\tau s}^* - Y_{\ell s} Y_{\tau d}^*$, $f_\pi
\to \sqrt{2} f_K$, and $\hat m_\pi \to \hat m_K$. For the decay channel
$\tau \to \mu \eta$ we include amplitudes for both $s\bar s$ and $d\bar d$
components of $\eta$ by replacing in~Eq.~\eqref{eq:tauwidth}
$\left|Y_{\ell d} Y_{\tau d}^*\right|^2 f_\pi^2 \to \left|f_\eta^q
  Y_{\mu d} Y_{\tau d}^*+\sqrt{2} f_\eta^s Y_{\mu s} Y_{\tau
    s}^*\right|^2$, where $f_\eta^{q,s}$ are the decay constants of
$\eta$ through $(\bar d \gamma^\mu \gamma_5 d + \bar u \gamma^\mu
\gamma_5 u)/\sqrt{2}$ and $\bar s \gamma^\mu \gamma_5 s$ operators,
respectively. Following~\cite{Feldmann:1998sh}, we include the effects of $\eta-\eta'$ mixing by using $f_\eta^q = f_q \cos \phi$ and $f_\eta^s = -f_s
\sin\phi$ with phenomenologically viable numerical values of $f_q
= 1.07 f_\pi$, $f_s = 1.34 f_\pi$, and $\phi = 39.3^\circ$. With
remaining numerical values $f_\pi = 130.4$\,MeV~\cite{Rosner:2010ak},
$f_K=156$\,MeV~\cite{Laiho:2009eu}, and the relevant $90$\,\% C.L. upper
bounds on the branching fractions~\cite{Nakamura:2010zzi} we find a
set of constraints shown in Tab.~\ref{tab:taudecays}.
\begin{table}[tb]
  \centering
  \begin{tabular}[c]{lcccr@{$\,\,<\,\,$}l}
    \hline
     decay mode && $90$\,\% C.L. exp. bound on $\mc B$&& \multicolumn{2}{c}{$1\,\sigma$ upper bound in units $(m_\Delta/400\e{GeV})^4$} \\\hline
$\tau \to e \pi^0$  && $8.0\E{-8}$  && $\left|Y_{ed} Y_{\tau
    d}^*\right|^2$&$1.9\E{-4}$\\
$\tau \to \mu \pi^0$ && $1.1\E{-7}$ && $\left|Y_{\mu d} Y_{\tau
    d}^*\right|^2$&$2.7\E{-4}$\\
$\tau \to e K_S$ && $3.3\E{-8}$ && $\left|Y_{ed} Y_{\tau s}^* - Y_{es} Y_{\tau d}^*\right|^2$&$3.2\E{-5}$\\
$\tau \to \mu K_S$ && $4.0\E{-8}$ && $\left|Y_{\mu d} Y_{\tau s}^* -
  Y_{\mu s} Y_{\tau d}^*\right|^2$&$4.0\E{-5}$\\
$\tau \to \mu \eta$ && $6.5\E{-8}$ && $\left|0.69\,Y_{\mu d} Y_{\tau
    d}^* - Y_{\mu s} Y_{\tau s}^*\right|^2$ &$1.3\E{-4}$\\\hline
  \end{tabular}
  \caption{Limits on $Y$ couplings coming from upper bounds on
    $\tau \to P \ell$ branching fractions, determined at the $B$-factories
    and compiled by~\cite{Nakamura:2010zzi}.}
  \label{tab:taudecays}
\end{table}

\subsubsection{$\mu-e$ conversion in nuclei}
Four fermion effective Lagrangian~\eqref{eq:Heff} contains also the LFV
terms $(\bar d \gamma_\mu P_R d) \{\bar \mu \gamma^\mu P_R e,  \bar e \gamma^\mu P_R \mu\}$. The most stringent bound on
such interactions is expected from experimental searches for
$\mu-e$ conversion in nuclei. In order to derive the relevant
constraints one needs to calculate the appropriate nuclear matrix
elements of the above operators.  A detailed analysis has been carried
out in~\cite{Kitano:2002mt}.  We can write the nuclear $\mu-e$
conversion rate as
\begin{equation}
\Gamma_{\rm conversion} =  \frac{|Y_{ed}Y_{\mu d}^*|^2}{4m_\Delta^4} |V^{(p)} + 2 V^{(n)}|^2\,,
\end{equation}
where the nuclear matrix elements $V^{(p,n)}$, calculated
in~\cite{Kitano:2002mt} for titanium and gold nuclei are given in
Tab.~\ref{table:nuclear}.
Presently the most stringent bounds on $\mc{B}_{\mu e} \equiv \Gamma_{\rm
  conversion}/ \Gamma_{\rm capture}$ was set by the SINDRUM
collaboration with $\mc{B}^{\rm (Ti)}_{\mu e} < 4.3 \times 10^{-12}$
\cite{Dohmen:1993mp} and $\mc{B}_{\mu e}^{\rm (Au)} < 7 \times 10^{-13}$
\cite{Bertl:2006up}, both at 90\,\% C.L.\,. Comparing these with our
theoretical expressions we obtain the corresponding $1\,\sigma$ bounds
\begin{equation}
  |Y_{ed}Y_{\mu d}^*|^2 < 1.9 (20)\E{-13} \left(\frac{m_\Delta}{400\e{GeV}}\right)^4 \quad {\rm from~Au(Ti)}\,.
\end{equation}
\begin{table}[tb]
  \begin{center}
    \begin{tabular}{lccc} \hline Nucleus & $V^{(p)} [m_\mu^{(5/2)}]$ &
      $V^{(n)} [m_\mu^{(5/2)}]$ & $\Gamma_{\rm capture}[10^6 s^{-1}]$
      \\ \hline
      ${\rm Ti}^{48}_{22}$    &              0.0396            &             0.0468              &  2.59 \\
      ${\rm Au}^{197}_{79}$ & 0.0974 & 0.146 & 13.07 \\ \hline
  \end{tabular}
  \end{center}
  \caption{Data taken from Tables I and VIII of~\cite{Kitano:2002mt}.}
  \label{table:nuclear}
\end{table}
Note that the same couplings also appear in the $\pi^0 \to e^\pm \mu^\mp$ decay
branching fraction, whose expectation is thus pushed far below the current experimental upper bound
of $\sim 10^{-10}$.

\subsection{One-loop effects of $\Delta$}
Next we turn our attention to observables which are affected by leptoquark couplings of $\Delta$ at the one-loop
level. These are $K-\bar K$ and
$B-\bar B$ mixing amplitudes,
LFV neutral current processes like the radiative $\mu$ and $\tau$
decays, as well as flavor diagonal observables, such as the anomalous magnetic moments of leptons or the decay width of the $Z$ to $b\bar b$ pairs. With the exploratory nature of our study in mind, we do not
consider nonlocal loop contributions due to the effective four-fermion
Lagrangian~\eqref{eq:Heff}, since such effects are constrained by the tree-level processes already considered in Sec.~\ref{sec:treelevel}. The particular case of new absorptive contributions affecting $B_s-\bar B_s$ oscillations will be discussed in Sec.~\ref{sec:fit}.

\subsubsection{$\epsilon_K$ and $\Delta m_K$}

\begin{figure}[h]
  \centering
  \begin{tabular}{cp{1cm}c}
    \includegraphics[width=0.2\textwidth]{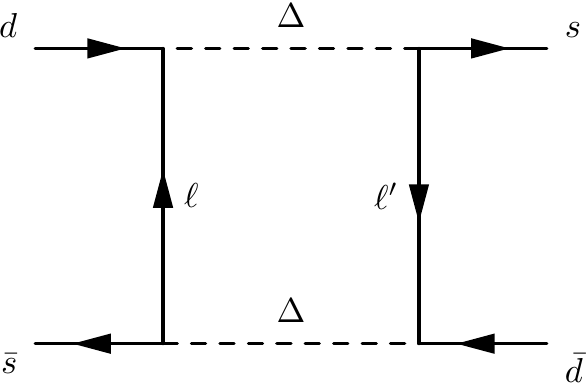} & &\includegraphics[width=0.2\textwidth]{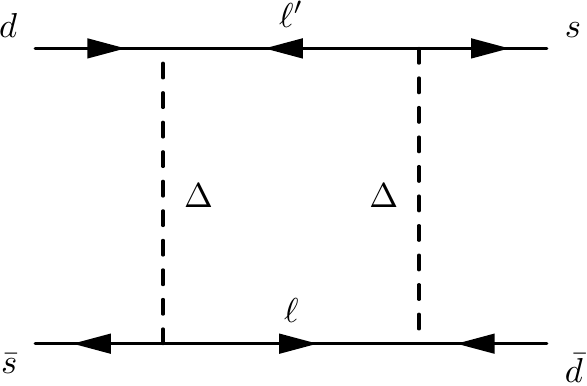}
  \end{tabular}
  \caption{$K-\bar K$ mixing diagrams with leptons and $\Delta$ in the box loop.}
  \label{fig:KKmixing}
\end{figure}
The SM result for the dispersive mixing matrix element, relevant
for $\epsilon_K$ is~\cite{Buras:1990fn}
\begin{align}
  \label{eq:9}
  M_{12K}^\mrm{SM} &= \frac{G_F^2 m_W^2}{12\pi^2} f_K^2 m_K \hat B_K
  \left[\lambda_c^{2} \eta_1 S_0(x_c) + \lambda_t^{2} \eta_2
    S_0(x_t)+2\lambda_c \lambda_t \eta_3 S_0(x_c,x_t)
  \right]\,.
\end{align}
$S_0$ is the Inami-Lim box loop function~\cite{Inami:1980fz} and factors
$\lambda_i = V_{is} V_{id}^*$ are the appropriate CKM weights. Explicit
$\lambda_u$ contributions are eliminated using the
Glashow-Iliopoulos-Maiani mechanism. Parameters $\eta_1$, $\eta_2$,
and $\eta_3$ account for the QCD renormalization effects and are known to NLO ($\eta_{1,2}$) or NNLO ($\eta_3$)
order. The decay constant $f_K$ and the reduced bag parameter $\hat
B_K$, both nonperturbative QCD parameters, are provided by lattice QCD
calculations. Values of all the relevant experimental as well as
theoretical parameters are compiled, together with their uncertainties, in
Tab.~\ref{tab:KKpars}.
\begin{table}[tb]
  \centering
  \begin{tabular}{|c|r@{.}lr|}
    \hline
    $|\epsilon_K|$ & $2$ & $228(11)\E{-3}$&\cite{Nakamura:2010zzi}\\
    $\Delta m_K$ & $3$ & $483(6)\E{-15}$\,GeV&\cite{Nakamura:2010zzi}\\
    $\phi_\epsilon$ &  $43$ & $5(7)^\circ$&\cite{Nakamura:2010zzi}\\\hline
    $f_K$ & $0$ & $1560(11)$\,GeV&\cite{Laiho:2009eu}\\
    $\hat B_K$ & $0$ & $725(26)$&\cite{Laiho:2009eu}\\ 
    $\kappa_\epsilon$ & $0$ & $94(2)$&\cite{Buras:2010pza}\\\hline
    $\eta_1$ & $1$ & $31({}^{+25}_{-22})$&\cite{Herrlich:1993yv} \\
    $\eta_2$ & $0$ & $57(1)$&\cite{Buras:1990fn,Buras:2011we}\\
    $\eta_3$ & $0$ & $496(47)$&\cite{Brod:2010mj}\\\hline
  \end{tabular}
  \caption{Experimental, nonperturbative, and perturbative parameters relevant for
    $\epsilon_K$ and $\Delta m_K$ observables.}
  \label{tab:KKpars}
\end{table}
The $K-\bar K$ transition is mediated also by box diagrams involving the
$\Delta$ and leptons, as shown in Fig.~\ref{fig:KKmixing}, that
generate an additional right-handed current operator in
the effective Hamiltonian~\cite{Dighe:2010nj}
\begin{equation}
  \label{eq:HeffKKbar}
  \mc{H}_{\Delta S=2}^{\Delta} = \frac{1}{128\pi^2 m_\Delta^2}
  \left[\sum_\ell Y_{\ell d} Y_{\ell s}^* \right]^2\, (\bar d_R
  \gamma^\mu s_R) (\bar d_R  \gamma_\mu s_R)\,.
\end{equation}
The dispersive mixing matrix element $M_{12K}$ induced by $\Delta$
is therefore
\begin{equation}
  \label{eq:11}
  M_{12K}^\Delta = \frac{1}{384\pi^2 m_\Delta^2} f_K^2 m_K \hat B_K
  \eta_2 \left[\sum_\ell Y_{\ell d} Y_{\ell s}^* \right]^2,
\end{equation}
where we have neglected the small QCD running effects from the
$\Delta$ mass scale to the EW scale and simply use $\eta_2$ to
describe the renormalization group evolution of $\Delta$ contributions
down to the hadronic scale.
The observable measuring the CP-even component of the $K_L$ mass-eigenstate,
$\epsilon_K$, is defined as the ratio of isospin singlet amplitudes of
$K_{S(L)} \to \pi\pi$ decays
\begin{equation}
 \epsilon_K \equiv \frac{A(K_L \to (\pi\pi)_{I=0})}{A(K_S \to (\pi\pi)_{I=0})}\,,
\end{equation}
and is related to the imaginary part of the dispersive mixing amplitude
as~\cite{Buras:2010pza}
\begin{equation}
 \epsilon_K = \kappa_\epsilon \frac{e^{i \phi_\epsilon}}{\sqrt{2}}
  \frac{\Im M_{12K}}{\Delta m_K}\,.
\end{equation}
Here $\Delta m_K$ is the measured mass difference between $K_L$ and
$K_S$ eigenstates, while $\phi_\epsilon$ is the superweak phase,
given by $\phi_\epsilon=\arctan(2 \Delta m_K/\Delta\Gamma_K)$. The
overall factor $\kappa_\epsilon$ contains long distance corrections
and uncertainties~\cite{Buras:2010pza}. The resulting constraint on
the $Y$ couplings is then
\begin{align}
  \label{eq:epskconst}
\Big|G_F^2 m_W^2 \Im\big[ \lambda_c^{2} \eta_1 S_0(x_c) + &\lambda_t^{2} \eta_2
    S_0(x_t)+2\lambda_c \lambda_t \eta_3 S_0(x_c,x_t) \big] +
  \frac{\eta_2}{16 m_\Delta^2} \Re [\sum_\ell Y_{\ell d}
    Y_{\ell s}^* ] \Im [\sum_\ell Y_{\ell d} Y_{\ell s}^* ]\Big|\nonumber\\
  &= \frac{12\sqrt{2}\pi^2}{f_K^2 \hat B_K
    \kappa_\epsilon} \frac{\Delta m_K}{m_K} |\epsilon_K| = 1.57(7)\E{-13}\e{GeV}^{-2}\,,
\end{align}
where on the right-hand side, we have combined the experimental and theoretical (hadronic)
uncertainties by summing them in squares. Nonetheless, some theoretical uncertainty coming from the QCD renormalization factors
$\eta_{1,2,3}$ still remains on the left-hand side. In the fit we allow them to freely vary within the intervals
determined by their theoretical uncertainties (see Tab.~\ref{tab:KKpars}).

The measured mass difference $\Delta m_K$, on the other hand, mostly
probes the real part of the mixing amplitude
$M_{12K}$~\cite{Buras:1998raa}. It receives potentially important
contributions from SM long distance dynamics leading to large
theoretical uncertainties in its
prediction~\cite{Herrlich:1993yv}. Therefore we conservatively assume
that the short distance contribution of $M_{12K}^{\Delta}$ must be
smaller than half the experimental value of $\Delta m_K$ at
$1\,\sigma$ C.L.:
\begin{equation}
  \label{eq:dmK}
  \Delta m_K^{\Delta} \simeq \Re\,M_{12K}^{\Delta} = \frac{1}{192\pi^2 m_\Delta^2} f_K^2 m_K \hat B_K
  \eta_2 \Re \left[\sum_\ell Y_{\ell d} Y_{\ell s}^* \right]^2
  < 1.74\E{-15}\e{GeV}\,.
\end{equation}
The conservative assumption for the bound~\eqref{eq:dmK} allows us to neglect
uncertainties of all the theoretical parameters and extract the following $1\,\sigma$ bound
on the real part of the $Y$ combination
\begin{equation}
  \label{eq:16}
  \Re \left[\sum_\ell Y_{\ell d} Y_{\ell s}^* \right]^2 < 1.1\E{-4} \left(\frac{m_\Delta}{400\e{GeV}}\right)^2\,.
\end{equation}

\subsubsection{$B_d-\bar B_d$ and $B_s-\bar B_s$ mixing}

The time evolution of the $B-\bar B$ system is described by the average
mass $m$, width $\Gamma$, and three mixing parameters
\begin{equation}
  \label{eq:2}
  |M_{12}|,\qquad |\Gamma_{12}|,\qquad \phi = -\arg(M_{12}/\Gamma_{12})\,.
\end{equation}
All five parameters can be identified by diagonalizing the effective
Hamiltonian
\begin{equation}
H_\mrm{eff} = M - \frac{i}{2} \Gamma
=
\begin{pmatrix}
  m & M_{12}\\
  M_{12}^* & m
\end{pmatrix}
-\frac{i}{2}
\begin{pmatrix}
  \Gamma & \Gamma_{12}\\
  \Gamma_{12}^* & \Gamma
\end{pmatrix}\,,
\end{equation}
whose off-diagonal elements are defined as $(H_{\rm eff})_{12} = \Braket{B |
  \mc{H}^{\Delta B=2}_\mrm{eff} + \rm{nonlocal \, interactions} | \bar B}/(2 m)$
and $\Gamma_{12}$ contains all on-shell contributions of intermediate
poles. Heavy and light ($H$ and $L$) mass-eigenstates are defined as
\begin{equation}
 \ket{B_{H,L}} = p \ket{B} \pm q \ket{\bar B}\,,
\end{equation}
and their eigenvalues are, in the appropriate limit $|\Gamma_{12}|
\ll |M_{12}|$, in turn connected to the measurements of $\Delta m$ and
$\Delta \Gamma$ as
\begin{subequations}
\begin{align}
  \label{eq:17}
  \Delta m &\equiv m_H-m_L = 2|M_{12}|\,,\\
  \Delta \Gamma &\equiv \Gamma_L - \Gamma_H =  2 |\Gamma_{12}| \cos\phi\,.
\end{align}
\end{subequations}
If CP violation in decays is negligible, one can
extract the phase $\phi$ from the semileptonic time-dependent CP asymmetry
\begin{equation}
  \label{eq:asldef}
  a_{\rm sl} (t) = \frac{\Gamma(\bar B(t) \to \ell^+ X)-\Gamma(B(t)
    \to \ell^- X)}{\Gamma(\bar B(t) \to \ell^+ X)+\Gamma(B(t) \to
    \ell^- X)}= \frac{\Delta \Gamma}{\Delta m} \tan \phi\ \,.
\end{equation}
The overall $\Delta \Gamma/\Delta m$ factor renders this asymmetry
very small.
Measurements of $a_{\rm sl}^{(d)}$ in the $B_d$ system have been performed
at the $B$-factories and a world average~\cite{Asner:2010qj} is
consistent with zero, albeit with much larger errors than the SM
predicted value. Direct measurement of $a_{\rm sl}^{(s)}$ is not
available, however D\O \, and CDF experiments~\cite{Asner:2010qj} have
measured the charge asymmetry of same-charge dimuon events coming from
inclusive $b$-decays, which is a linear combination of $a_{\rm sl}^{(d)}$
and $a_{\rm sl}^{(s)}$, thus allowing one to extract $a_{\rm sl}^{(s)}$.
Especially the D\O \, measurements~\cite{Abazov:2011yk,Abazov:2010hv,:2008fj} point at
an unexpectedly large mixing phase $\phi$ and exclude the SM value of
$a_{\rm sl}^{(s)}$ with more than $3\,\sigma$
significance~\cite{Ligeti:2010ia,Bauer:2010dga,Lenz:2010gu}.

More effectively, one can extract the phase in the dispersive mixing amplitude
$M_{12}$ from the time-dependent CP asymmetry in decays of $B$~$(\bar B)$
to CP eigenstates
\begin{equation}
  \label{eq:CPeigenstates}
  A_{\rm CP}^f (t)= \frac{\Gamma(B(t) \to f) - \Gamma(\bar B(t) \to
    f)}{\Gamma(B(t) \to f) +  \Gamma(\bar B(t) \to
    f)} = \eta_f \Im \left(\frac{p}{q}\right) \sin \Delta m\, t\,,
\end{equation}
where we have assumed a tree-level dominated decay mechanism, 
negligible CP violation in the decay, and also $|p/q|=1$. CP parity of
the final state is denoted as $\eta_f$. To leading order in
$|\Gamma_{12}/M_{12}|$, $\Gamma_{12}$ cancels out and one is sensitive
to the phase of $M_{12}$ through $\Im (p/q) = \Im
(M_{12})/|M_{12}|$. This phase is interpreted within the SM as an angle of
the unitarity triangle $\sin 2\beta$~$(\sin 2\beta_s)$ in the case of the
$B_d$~$(B_s)$ system. Note that the weak phase of the absorptive part
is negligible in the SM and also difficult to enhance in most NP
scenarios and thus $\phi_{d(s)} \approx -2\beta_{(s)}$ (we will
comment on the recent study~\cite{Dighe:2010nj} of $\Delta$
contributions to the absorptive amplitude in Sec.~\ref{sec:fit}).

At this point we shall include as experimental
constraints only the measurements of $\sin 2\beta$ and the mass splittings
$\Delta m_d$, $\Delta m_s$. We will address the allowed ranges of $\phi_s$ in the fit part in Sec.~\ref{sec:fit}.
\begin{table}[!htb]
  \centering
  \begin{tabular}{|c|r@{.}lr|}
    \hline
    $\sin 2\beta$ & $0$ & $673(23)$&\cite{Asner:2010qj}\\
    $\Delta m_d$ &  $0$& $507(5)\e{ps}^{-1}$&\cite{Nakamura:2010zzi}\\
    $\Delta m_s$ & $17$&$77(12)\e{ps}^{-1}$&\cite{Nakamura:2010zzi}\\\hline
    $f_{B_s}(\hat B_{B_s})^{1/2}$ & $0$ & $275(15)\e{GeV}$&\cite{Laiho:2009eu}\\ 
    $\xi$ & $1$ & $237(32)$&\cite{Laiho:2009eu}\\\hline
    $\eta_B$ & $0$ & $55(1)$&\cite{Buras:1990fn,Urban:1997gw} \\\hline
  \end{tabular}
  \caption{Experimental, nonperturbative, and perturbative parameters relevant for
    $B_{d(s)}-\bar B_{d(s)}$ mixing.}
  \label{tab:BBpars}
\end{table}
The SM prediction for the dispersive matrix element $M_{12}$  in $B_d$ mixing (with
obvious replacements in the case of $B_s$ mixing), is dominated by the
short distance box diagrams involving the top quark
\begin{equation}
  \label{eq:M12BBbar}
    M_{12B}^\mrm{SM} = \frac{G_F^2 m_W^2}{12\pi^2} f_B^2 m_B \hat B_B
    (V_{tb} V_{td}^*)^2 \eta_B
    S_0(x_t)\,.
\end{equation}
The three theoretical parameters here are again the perturbative QCD renormalization factor $\eta_B$, and the nonperturbative hadronic
parameters $f_B$ and $\hat B_B$.
Box diagrams with $\Delta$, analogous to the ones of $K-\bar K$
in Fig.~\ref{fig:KKmixing}, can shift the value of
$M_{12}^\mrm{SM}$ by
\begin{equation}
  \label{eq:5}
  M_{12B}^{\Delta} = \frac{1}{384\pi^2 m_\Delta^2} f_B^2 m_B \hat B_B
  \eta_B \left[\sum_\ell Y_{\ell d} Y_{\ell b}^* \right]^2\,,
\end{equation}
where we again neglect the difference between the $\Delta$ and EW matching scales. Instead of using $\Delta m_d$ and $\Delta m_s$ as individual fit constraints, we opt to
trade $\Delta m_d$ for the ratio $\Delta m_s/\Delta m_d$ depending
on the hadronic parameter $\xi (\equiv \hat{B}_{B_s}
\sqrt{f_{B_s}}/\hat{B}_{B_d} \sqrt{f_{B_d}})$ that can be determined
 reliably using lattice QCD techniques
\begin{subequations}
\label{eq:dmB}
\begin{align}
  \label{eq:msmd}
  &\left|\frac{(V_{tb}V_{ts}^{*})^{2} S_0(x_t) +
      (32 m_\Delta^2 G_F^2 m_W^2)^{-1} \left(\sum_\ell Y_{\ell s}
        Y_{\ell b}^*\right)^2}{(V_{tb}V_{td}^{*})^{2} S_0(x_t) +
      (32 m_\Delta^2 G_F^2 m_W^2)^{-1} \left(\sum_\ell Y_{\ell d}
        Y_{\ell b}^*\right)^2}\right| = \frac{\Delta m_s}{\Delta
    m_d}\, \frac{m_{B_d}}{m_{B_s}}\,\xi^{-2} = 22.5(12)\,,\\
  \label{eq:dms}
   &\left|(V_{tb}V_{ts}^{*})^{2} S_0(x_t) +  (32 m_\Delta^2 G_F^2 m_W^2)^{-1}
  \left[\sum_\ell Y_{\ell s} Y_{\ell b}^*
  \right]^2 \right| = \Delta m_s\,\frac {6\pi^2}{G_F^2 m_W^2 \eta_B \hat B_{B_s}
  f_{B_s}^2 m_{B_s} } = 3.53(39)\E{-3}\,.
\end{align}
\end{subequations}
On the right-hand sides, we have combined the experimental and
theoretical errors in quadrature. The relevant numerical inputs
are compiled in Tab.~\ref{tab:BBpars}.

On the other hand, in the $\sin 2\beta$ constraint all dependence on
theoretical (in particular hadronic) parameters drops out
\begin{equation}
  \label{eq:sin2betaAlt}
   \frac{\Im\left[
   S_0(x_t)  (V_{tb}V_{td}^{*})^{2} + (32 m_\Delta^2 G_F^2 m_W^2)^{-1}  \left(\sum_\ell Y_{\ell d}
        Y_{\ell b}^*\right)^2 \right]}
{ \left| S_0(x_t)  (V_{tb}V_{td}^{*})^{2} + (32m_\Delta^2 G_F^2 m_W^2)^{-1}  \left(\sum_\ell Y_{\ell d}
        Y_{\ell b}^*\right)^2 \right|} = \sin 2\beta\,.
\end{equation}

\subsubsection{Anomalous magnetic and electric dipole moments}

The electromagnetic interactions of an on-shell fermion can be parameterized in terms of
parity conserving and parity violating form
factors~\cite{Jegerlehner:2009ry}
\begin{subequations}
\begin{align}
  \mc{A}^\mu &\equiv -ie \bar{u}(p',s') \Gamma^\mu  u(p,s),\\
  \Gamma^\mu &= F_1 \gamma^\mu + \frac{F_2}{2m_\mu}
    i\sigma^{\mu\nu}q_\nu + F_3 \sigma^{\mu\nu}q_\nu \gamma_5 +F_4 (2m
    q^\mu + q^2 \gamma^\mu)\gamma_5\label{eq:ffs}\,,
\end{align}
\end{subequations}
where $q = p - p'$. This is the most general form of the photon off-shell
amplitude obeying the Ward identity of quantum
electrodynamics
\begin{equation}
q_\mu \mc{A}^\mu  = 0\,.
\end{equation}
Renormalized charge of a muon is $-e$ and so $F_1(0) = 1$
exactly. A finite $F_3(0)$ would signal a nonzero electric dipole moment in presence of CP violating phases in the renormalized vertex. $F_4(0)$ is called the anapole moment.  The form factor
$F_2$, which is the source of the anomalous magnetic moment, enters in
the gyromagnetic ratio as $g = 2(F_1(0)+F_2(0))$.  Comparing precise
measurements of these form factors against theoretical
higher-order predictions presents powerful tests of the SM and its
extensions. In the recent years, the experimental result on the anomalous magnetic moment of the muon $a_\mu \equiv (g-2)_\mu/2$  from
BNL~\cite{Bennett:2004pv} has been about $3\,\sigma$ above theoretical
predictions within the SM~\cite{Jegerlehner:2007xe}
\begin{subequations}
\begin{align}
 a_\mu^\mrm{exp} &= 1.16592080(63)\E{-3}\,,\\
 a_\mu^\mrm{SM} &= 1.16591793(68)\E{-3}\,.
\end{align}
\end{subequations}
Treating both experimental and theoretical uncertainties as Gaussian, we
may identify the missing contribution to $a_\mu$
\begin{equation}
\delta a_\mu = a_\mu^\mrm{exp} - a_\mu^\mrm{SM} = (2.87\pm 0.93)\E{-9}\,,
\end{equation}
with the presence of NP.
\begin{figure}[h]
  \centering
  \begin{tabular}{lcccr}
    \includegraphics[width=.27\textwidth]{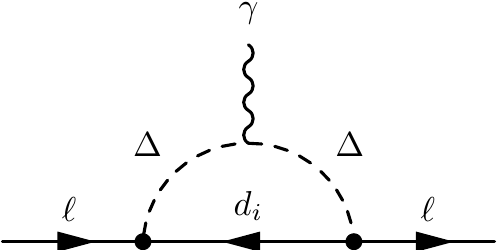} &&&&   \includegraphics[width=.27\textwidth]{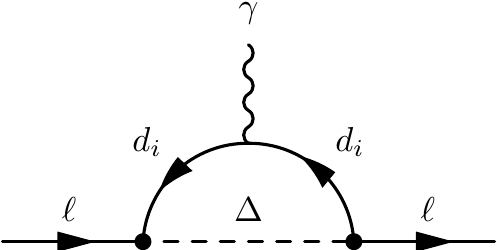} 
  \end{tabular}
  \caption{Diagrams with $\Delta$ and down-quarks contributing to
    the lepton anomalous magnetic moments.}
  \label{fig:aell}
\end{figure}
The leading $\Delta$ contributions to $a_\mu$ with $\Delta$ and down quarks $d_i$ running in the
loop~(Fig.~\ref{fig:aell}) are expected to be of the order $\sim
1/(4\pi)^2\,m_\mu^2/m_\Delta^2\, e|Y_{\mu i}|^2$ and have been previously computed in~\cite{Chakraverty:2001yg}.  We reproduce the
magnitude of $a_\mu$ of~\cite{Chakraverty:2001yg}, however with
an opposite overall sign
\begin{equation}
  a_\mu = \frac{3 m_\mu^2}{16\pi^2 m_{\Delta}^2} \sum_{i=d,s,b} |Y_{\mu i}|^2 \left[Q_\Delta f_\Delta (x_i) + Q_d f_d (x_i)\right],\qquad x_i=m_{d_i}^2/m_{\Delta}^2.
\end{equation}
Here the charges are $Q_{\Delta,d} = 4/3,\,-1/3$ while $f_{\Delta,d}$
are the loop functions
\begin{subequations}
\begin{align}
  f_\Delta(x) &= \frac{2 x^3 + 3 x^2 - 6 x^2 \log x -6 x+1}{6(x-1)^4}\,,\\
  f_d(x) &= \frac{-x^3+6 x^2-6 x \log x -3 x-2}{6(x-1)^4}\,.
\end{align}
\end{subequations}
In the limit $x_i \to 0$ the result becomes
\begin{equation}
a_\mu = \frac{3  m_\mu^2 (Q_\Delta-2Q_d)}{96 \pi^2 m_\Delta^2} \sum_{i=d,s,b} |Y_{\mu i}|^2 = \frac{1}{16\pi^2}\,\frac{m_\mu^2}{m_{\Delta}^2}\,\sum_{i=d,s,b} |Y_{\mu i}|^2.
\end{equation}
If we now saturate $\delta a_\mu$ with $a_\mu$ we find that 
a non-zero magnitude is preferred for a combination of the second row elements of $Y$
\begin{equation}
\label{eq:amu}
 \sum_{i=d,s,b} |Y_{\mu i}|^2 = (4.53 \pm 1.47)
 \E{-7}\times\frac{m_{\Delta}^2}{m_\mu^2} = (6.45\pm 2.09)\times \frac{m_\Delta^2}{(400\e{GeV})^2}\,.
\end{equation}
We will further explore the possible correlations of such effects with other constraints in Sec.~\ref{sec:fit}.

On the other hand, applying expression~\eqref{eq:amu} to the electron
case and requiring that $a^\Delta_e$ be smaller than the experimental
uncertainty, we find a $1\,\sigma$ bound on the first row of $Y$
\begin{equation}
 \sum_{i=d,s,b}|Y_{ei}|^2 < 8.8
 \E{-11}\times\frac{m_{\Delta}^2}{m_e^2} = 54 \times \frac{m_\Delta^2}{(400\e{GeV})^2}\,,
\end{equation}
where we have used the experimental uncertainty estimate of $\sigma_{a^\mrm{exp}_e} =
2.8\E{-13}$~\cite{Nakamura:2010zzi}.

Finally, we note that due to the Hermitian structure of $Y$ contributions to the EM interactions of quarks and leptons, no electric (or chromoelectric) dipole moments of either quarks or leptons are generated at the one loop level, regardless of the phases present in $Y$. Furthermore, even at the two loop level, non-zero contributions can only originate from mixed $W-\Delta$ loops. However, since $\Delta$ interactions are purely right-handed, such contributions are necessarily suppressed both by CKM factors and by insertions of the light quark or lepton masses. Therefore we do not consider them further.

\subsubsection{Flavor violating radiative decays}

The computation of $\Delta$ contributions to the LFV radiative muon decay is analogous to the magnetic
moment diagrams in Fig.~\ref{fig:aell} and results in the effective
Lagrangian
\begin{equation}
  \label{eq:12}
  \mc{L}^{\Delta}_{\mu \to e \gamma} = \frac{e}{64\pi^2
    m_\Delta^2}  \Bigg[\sum_{i=d,s,b} Y_{e i} Y_{\mu i}^*\Bigg]
  \bar e (\sigma^{\mu\nu} F_{\mu\nu}) (m_\mu P_L + m_e P_R) \mu\,.
\end{equation}
The decay width of $\mu \to e \gamma$ is then given by
\begin{equation}
  \label{eq:meg}
  \Gamma_{\mu \to e \gamma} = \frac{\alpha m_\mu^5 }{4096 \pi^4 m_\Delta^4}
  \Big|\sum_{i=d,s,b} Y_{e i} Y_{\mu i}^*\Big|^2\,. 
\end{equation}
The above expression can also be applied to the LFV decays of the $\tau$,
with obvious replacements in $Y$ indices and masses. 
Inequalities following from upper limits on branching fractions of
$\ell \to \ell' \gamma$ are shown in
Tab.~\ref{tab:meg}.
\begin{table}[tb]  \centering
  \begin{tabular}[c]{ccccr@{$\,\,<\,\,$}l}
    \hline
     decay mode && $90$\,\% C.L. exp.
    bound on $\mc B$&& \multicolumn{2}{c}{$1\,\sigma$ upper bound in units $(m_\Delta/400\e{GeV})^4$} \\\hline
$\mu \to e \gamma$  && $2.4\E{-12}$  && $|\sum_{i=d,s,b} Y_{e i}
  Y_{\mu i}^*|^2$&$4.6\E{-8}$ \\
$\tau \to \mu \gamma$ && $4.4\E{-8}$ && $|\sum_{i=d,s,b} Y_{\mu i} Y_{\tau i}^*|^2$&$4.8\E{-3}$ \\
$\tau \to e \gamma$ && $3.3\E{-8}$ && $|\sum_{i=d,s,b} Y_{e i} Y_{\tau i}^*|^2$&$3.6\E{-3}$\\\hline
 \end{tabular}
  \caption{Limits on $Y$ couplings coming from upper bounds of
    LFV radiative lepton decay branching fractions, taken from~\cite{Nakamura:2010zzi,Adam:2011ch}.}
  \label{tab:meg}
\end{table}
Consequently, measurements of these flagship LFV processes impose strict
requirements on the structure of $Y$, namely they require that rows of $Y$
are approximately orthogonal.  

On the other hand, the analogous constraints coming from the quark sector radiative decays are much weaker. The prominent example of $b\to s\gamma$ has recently been analyzed in~\cite{Bauer:2010dga}, where it was found that this decay is not very sensitive to the relevant $\Delta$ interactions, which contribute at one loop through the insertion of the Lagrangian~\eqref{eq:Heff}. In particular, the $\Delta m_s$ constraint (\ref{eq:dms}) yields consistently stronger bounds on the same combination of $Y$ elements for the experimentally allowed range of $\Delta$ masses.

\subsubsection{Decay width of $Z\to b \bar{b}$}
\begin{figure}[h]
  \centering
  \begin{tabular}{lcr}
    \includegraphics[width=.25\textwidth]{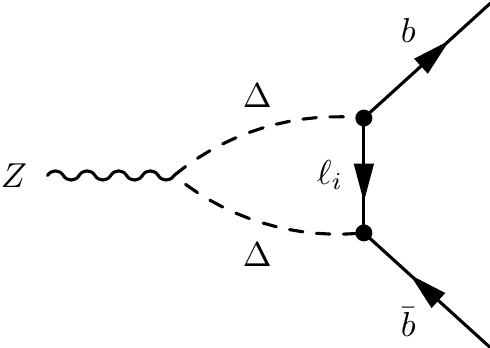} &&   \includegraphics[width=.25\textwidth]{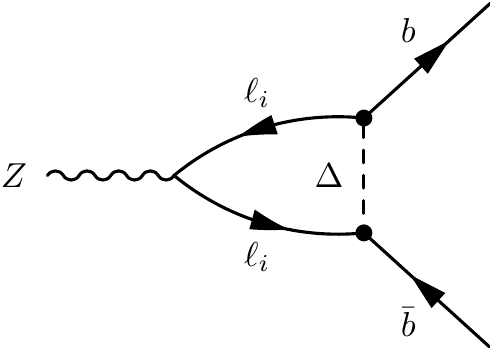} 
  \end{tabular}
  \caption{Diagrams with $\Delta$ and leptons modifying the $Zb\bar b$
    vertex.}
  \label{fig:Zbb}
\end{figure}
The experiments running on the LEP 1 collider performed precise measurements of the relative widths of $Z\to b \bar b$
and $Z \to $\,hadrons. In particular the experimental value of
\begin{equation}
R_b = \frac{\Gamma(Z\to b\bar{b})}{\Gamma(Z\to \mrm{hadrons})}\,,
\end{equation}
is according to the PDG~\cite{Nakamura:2010zzi} in good agreement with SM predicted value
\begin{subequations}
\begin{align}
R^\mrm{exp}_b &=0.21629(66)\,, \\
R^\mrm{SM}_b &=0.21578(5)\,.
\end{align}
\end{subequations}
The SM tree-level amplitude for the $Zb \bar b$ vertex is
\begin{eqnarray}
\mc{A}^{\mathrm{tree}}
&=&\mathrm{i} g_Z\Big[g_R^0 \mc{A}_R + g_L^0 \mc{A}_L \Big]\,,\qquad  \mc{A}_{L(R)} = Z_\mu\, \bar b \gamma_\mu P_{L(R)} b\,,\\
g_R^0 &=& \frac{1}{3}\sin^2\theta_W\,,\hspace{0.5cm} g_L^0 = -1/2 + \frac{1}{3}\sin^2\theta_W\,,\nonumber
\end{eqnarray}
where $g_Z = g/\cos \theta_W$. New contributions of $\Delta$ change $g_R^0$ to  $g_R = g_R^0 + \delta
g_R$, where
\begin{eqnarray}
\label{eq:dgr}
\delta g_R &=& \sin^2\theta_W
\frac{\sum_{\ell}|Y_{\ell b}|^2}{(4\pi)^2}\frac{1}{6x_Z^2}\Bigg[
\frac{17}{2}x_Z^2-2x_Z+\log(x_Z)\big(3x_Z^2-6x_Z+6\log(1+x_Z)\big)\\
&&-8f_1 +4f_2(2x_Z-x_Z^2)+6\mathrm{Li}_2(-x_Z) - \mathrm{i} \pi  \Big( 3x_Z^2 -6x_Z+6\log(1+x_Z)\Big) 
\Bigg]\,,\nonumber
\end{eqnarray}
and $x_Z = m_Z^2/m_\Delta^2$. Calculational details along with
functions $f_1$ and $f_2$ are given in the Appendix~\ref{app:Rb}.
Taking into account higher order SM corrections, the relative shift in $R_b$ due to such NP contributions can be written
as~\cite{Oakes:1999zi}
\begin{equation}
\label{eq:Rbchange}
\delta R_b = 2 R_b^{\mathrm{SM}}(1-R_b^{\mathrm{SM}})\frac{g_L^0
  \,\Re (\delta g_L) + g_R^0 \,\Re (\delta g_R)}{(g_L^0)^2+(g_R^0)^2}\,.
\end{equation}
One can check that the shift $\Re(\delta g_R)$ given in
Eq.~\eqref{eq:dgr} is negative and $\delta g_L = 0$ and consequently any
contributions of $\Delta$ necessarily worsen the agreement between theory and
experiment. If the discrepancy should be smaller than $1\,\sigma$
the following constraint has to be met
\begin{equation}
 \sum_\ell |Y_{\ell b}|^2 < 5.60 \left(\frac{m_\Delta}{400\e{GeV}}\right)^2
  +6.73 \left(\frac{m_\Delta}{400\e{GeV}}\right)+2.02\,.
\end{equation}
In derivation of this bound we have approximated $\delta g_R$ by a
polynomial in variable $m_\Delta$ and employed $\sin^2 \theta_W =
0.231$~\cite{Nakamura:2010zzi}.

On the other hand, the forward-backward asymmetry in $b\bar b$ production as measured at LEP exhibits a $2.7\,\sigma$ tension with the SM EW fit. Since $\Delta$ contributions only affect the right-handed effective $Zb\bar b$ coupling ($\delta g_R$) they cannot fully reconcile this tension~\cite{Oakes:1999zi} and we do not include this observable in the fit.

%
\section{Global fit of the leptoquark couplings}
\label{sec:fit}
%

In this section we perform a global fit of $Y$ to all the observables
listed in Sec.~\ref{sec:pheno}, while we keep fixed $m_\Delta =
400$\,GeV. We resort to a $\chi^2 (Y)$ statistic that we minimize to
find the point $\chi^2_\mrm{min} = \chi^2(Y_\mrm{best})$, where $Y$ is
by definition in best agreement with all the constraints. $\chi^2(Y)$
is written as a sum of Gaussian contributions of observables
$\mc{O}_i$
\begin{equation}
  \label{eq:chisq}
  \chi^2 (Y)= \sum_i \frac{\left(\mc{O}_i^\mrm{exp}-\mc{O}^\mrm{prediction}_i(Y)\right)^2}{\left(\sigma_i^\mrm{eff}\right)^2}\,.
\end{equation}
Values of $\mc{O}_i^\mrm{exp}$ and $\sigma_i^\mrm{eff}$ are central
values and errors, read-off from right-hand sides of constraining
equations in the preceding sections, whereas
$\mc{O}_i^\mrm{prediction}$ are the corresponding predictions in terms
of $Y_{ij}$, i.e., the left-hand sides of constraints, in the language
of Sec.~\ref{sec:pheno}. Majority of $\mc{O}_i^\mrm{exp}$ are upper
bounds which are modeled with $\chi^2$ centered at zero. This is
achieved in Eq.~\eqref{eq:chisq} by setting $\mc{O}_i^\mrm{exp} = 0$
and $\sigma_i^\mrm{eff}$ to the derived $68$\,\% C.L. upper
bound. Although not explicitly shown here, the $\chi^2$
function~\eqref{eq:chisq} depends also on the 4 Wolfenstein parameters
of the CKM matrix (they are present in meson mixing constraints) and we
treat them on the same footing as $Y$. We add to~\eqref{eq:chisq} a
Gaussian chi-square term which guides the CKM parameters to follow
probability distribution of Eqs.~\eqref{eq:CKMcv} and \eqref{eq:corr}.

Statistical interpretation of the value of $\chi^2$, i.e. the goodness of
fit, is performed using the standard $\chi^2$ probability distribution with
appropriate number of degrees of freedom, $N_\mrm{DOF} =
N_\mrm{observables} - N_\mrm{parameters} = 36-21$.
To find the allowed range of a single matrix element $|Y_{ij}|$, its
phase, or a function of several $Y$ elements, denoted in the following
generically as $z(Y)$, we minimize $\chi^2$ with $z(Y)$ fixed to some
chosen value $z_0$. Then all values $z_0$, where
\begin{equation}
  \label{eq:19}
  \mrm{min} \big[\chi^2 (Y)_{z(Y)=z_0}\big] - \chi^2_\mrm{min}< 1\,(4)
\end{equation}
form the $68.3\,(95.45)$\,\% C.L. interval for the parameter $z$. To
find confidence level regions in two-dimensional scans (with two fixed
quantities $z(Y)$, $w(Y)$) we utilize the $\chi^2$-distribution with 2
degrees of freedom. The difference $\Delta \chi^2(z_0, w_0) = \min
[\chi^2_{z(Y)=z_0, w(Y) = w_0}]-\chi^2_\mrm{min}$ for
points $(z_0, w_0)$ in the $N\,\sigma$ C.L. region in this case is
\begin{equation}
  \label{eq:21}
  1-\exp \left[-\Delta\chi^2(z_0,w_0)/2\right] < \mrm{erf} \left(\frac{N}{\sqrt{2}}\right)\,.
\end{equation}

\subsection{Structure of $Y$}
In a trivial case, when we set $Y=0$ to recover the SM, we find
$\chi^2_\mrm{min} = 12.5=9.5_{a_\mu} + 1.5_\mrm{CKM} + 0.8_{\Delta
  m_s} + \cdots$ with a dominant contribution from the $a_\mu$
anomaly.
If we let $Y$ take any value we find a global minimum
$\chi^2_\mrm{min} = 2.5 = 1.8_\mrm{CKM}+0.4_{\Delta m_s} + \cdots$ for $15$ degrees of freedom, which signals a very
good agreement of all predictions with the considered observables. In
particular, the best point perfectly resolves the anomalous magnetic
moment constraint $a_\mu$ and slightly improves quark flavor
constraints. The allowed $1$ and $2\,\sigma$ ranges of $Y$ matrix
elements are shown below
\begin{subequations}
\label{eq:Yranges}
\begin{eqnarray}
|Y^\mrm{(1\,\sigma)}| &\in&
\begin{pmatrix}
<1.4\E{-6} & <8.7\E{-5} & < 4.2\E{-4}\\
<3.6\E{-3} \cup [2.1,2.9]  & <3.6\E{-3} \cup [2.1, 2.9] & <6.2\E{-4} \cup [2.3,2.7]\\
<5.6\E{-3} & <8.1\E{-3} & <9.6\E{-3}
\end{pmatrix}\,,\\\nonumber\\
|Y^\mrm{(2\,\sigma)}| &\in&
\begin{pmatrix}
<2.2\E{-6} & < 1.4\E{-4} & <6.6\E{-4}\\
<5.6\E{-3} \cup [1.5,3.3] & <5.6\E{-3} \cup [1.5,3.3] & <9.7\E{-4} \cup [1.6,3.2]\\
<8.9\E{-3} & <1.4\E{-2} & <1.5\E{-2}
\end{pmatrix}\,.
\end{eqnarray}
\end{subequations}
\begin{figure}[!htb]
  \centering\begin{tabular}{ccc}
    \includegraphics[width=0.47\textwidth]{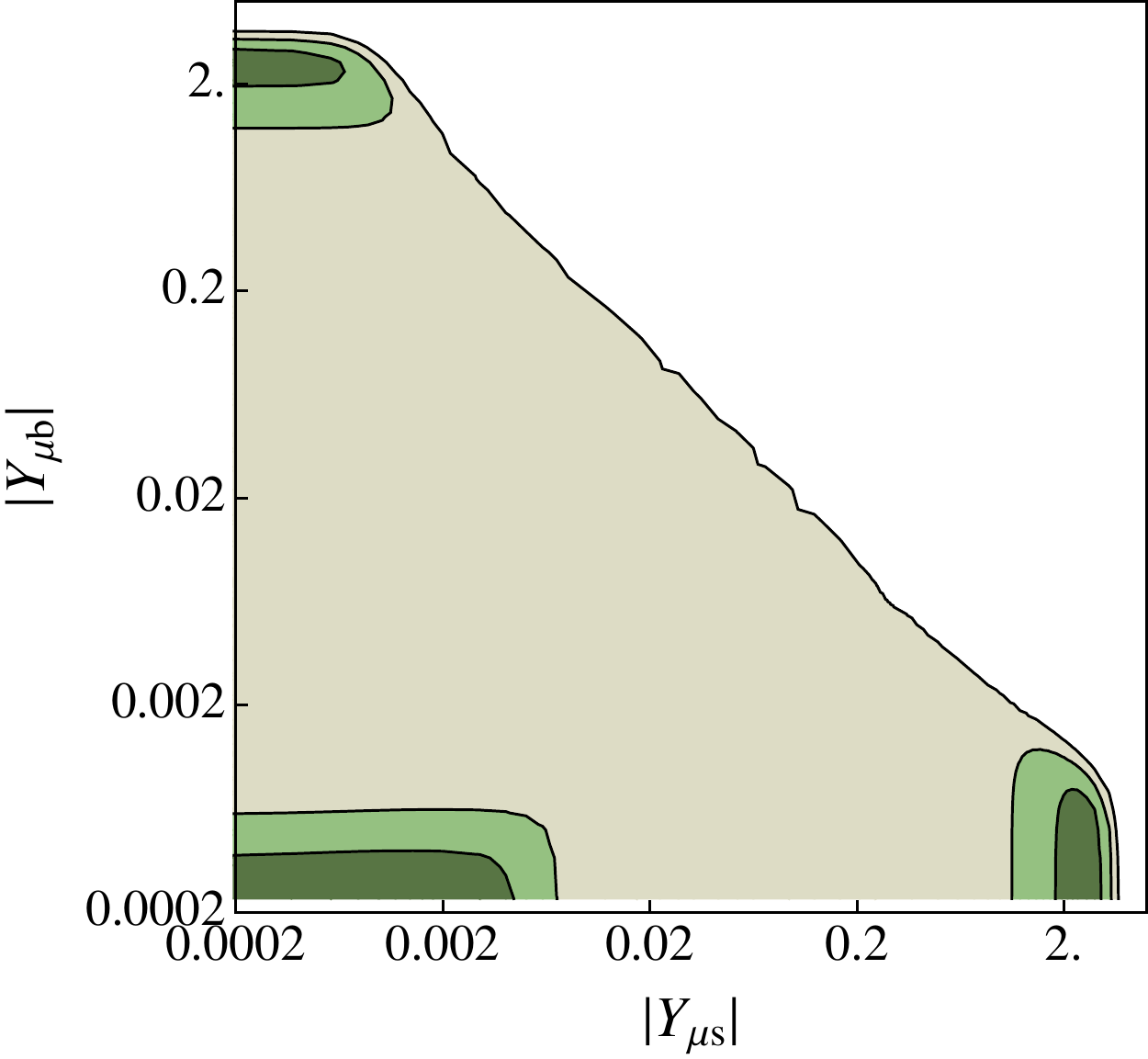} &&  \includegraphics[width=0.46\textwidth]{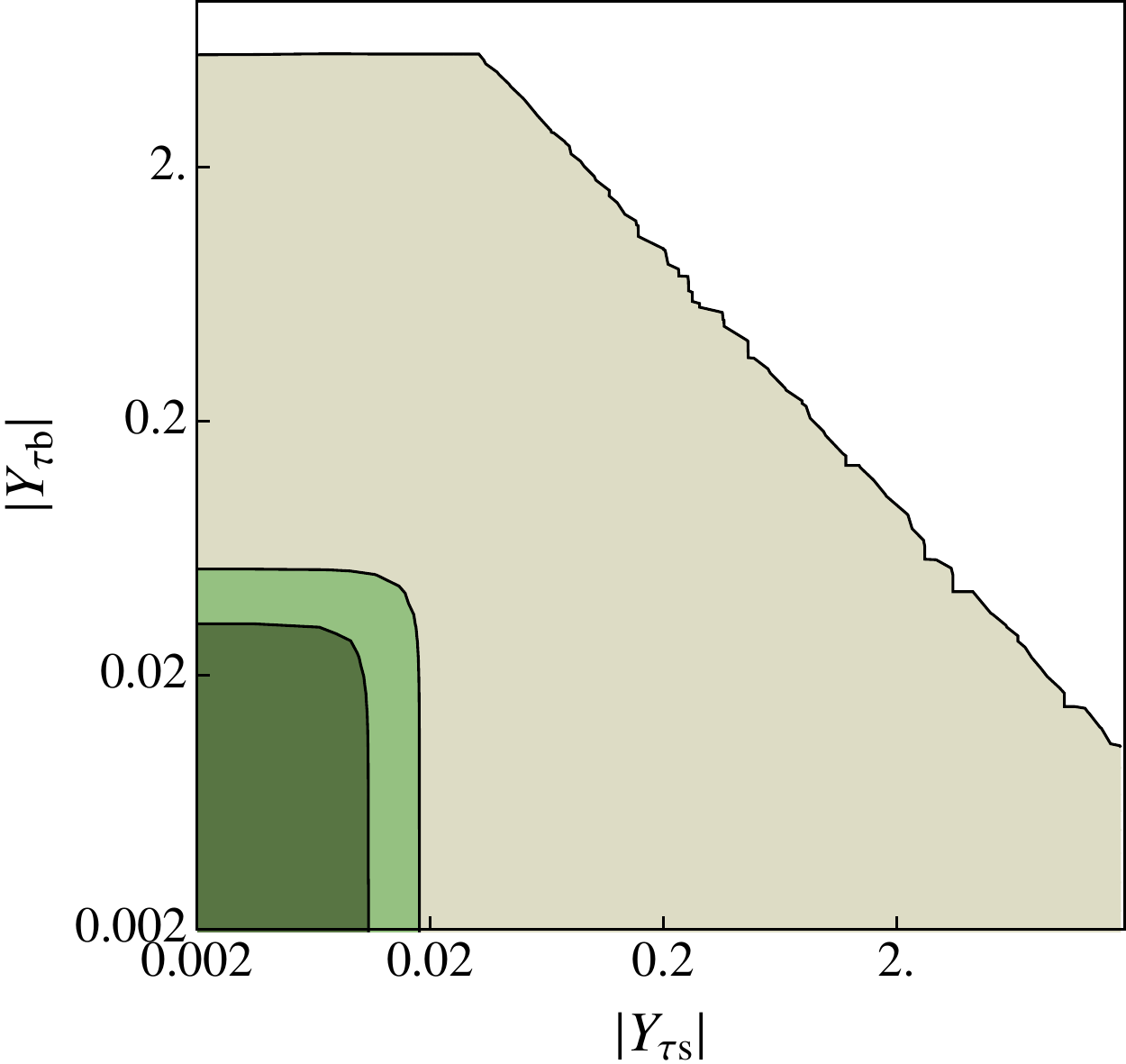}
    \end{tabular}
    \caption{Correlation between elements in the second ($|Y_{\mu s}|$
      and $|Y_{\mu b}|$) and third ($|Y_{\tau s}|$ and $|Y_{\tau b}|$)
      row. Dark green region is the $1\,\sigma$ contour, while the two
      lighter green contours are $2$ and $3\,\sigma$, respectively.}
  \label{fig:Yx2Yx3}
\end{figure}
\begin{figure}[!htb]
  \centering
  \includegraphics[width=0.45\textwidth]{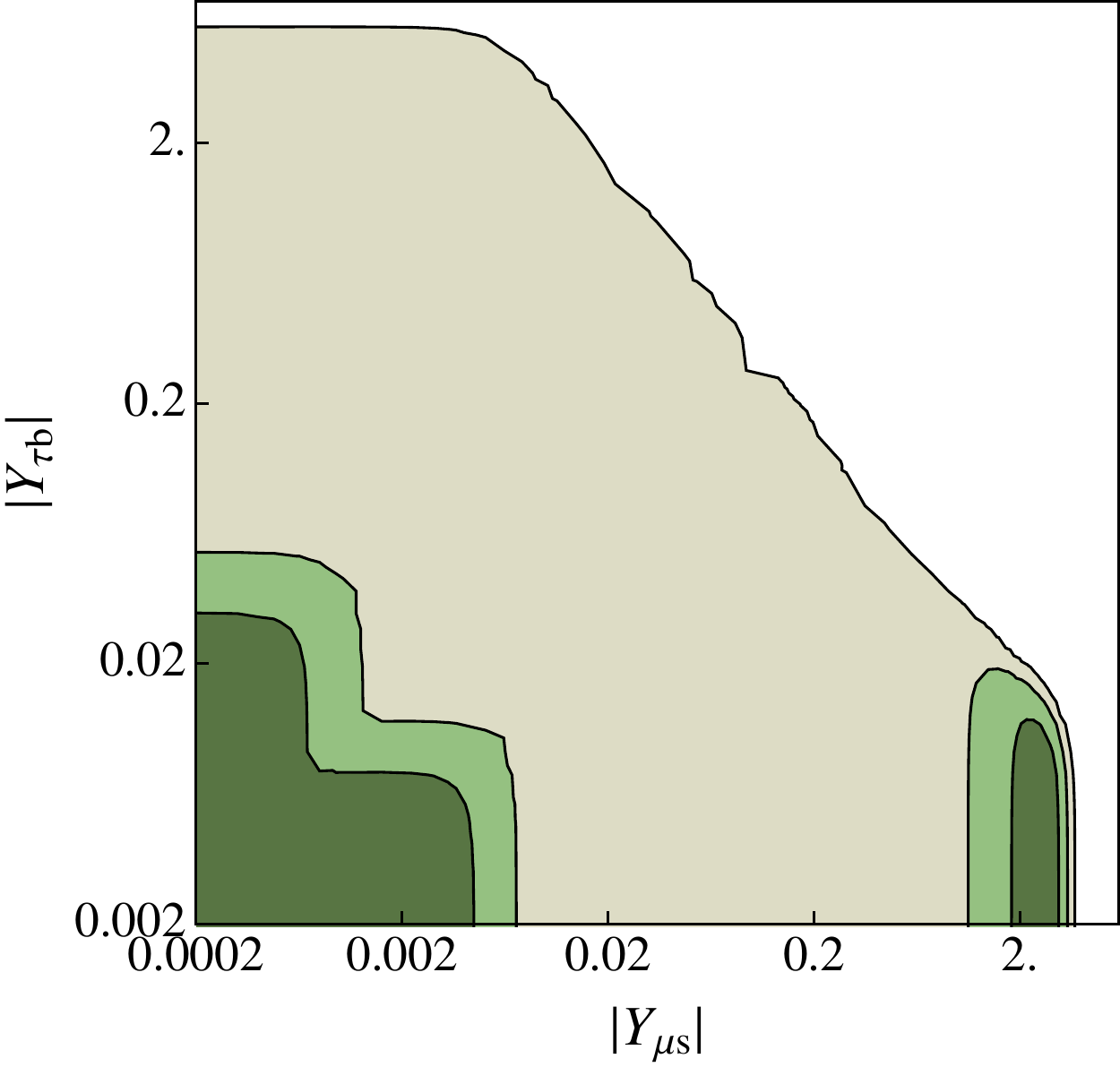}
  \caption{Correlation between diagonal $Y$ elements in the muon and
    tau rows. Dark green region is the $1\,\sigma$ contour, while the
    two lighter ones are $2$ and $3\,\sigma$, respectively.}
  \label{fig:diags}
\end{figure}
Couplings to the electron are strongly suppressed, while couplings to
the muon (the second row of $Y$) can take values of order $1$, in order to satisfy the
$a_\mu$ constraint. In the last row, elements $Y_{\tau s}$ and
$Y_{\tau b}$ can also be of order $0.01$ at $1\,\sigma$ C.L.\, We find some
interesting correlations between the second and third row elements,
shown in Figs.~\ref{fig:Yx2Yx3} and \ref{fig:diags}.

We find three distinct regimes in the second and third
row~(Figs.~\ref{fig:Yx2Yx3} and \ref{fig:diags}), depending on which
element in the second row is large. Pictorially, these hierarchies are
possible
\begin{equation}
  \label{eq:Yhierarchy}
  \begin{pmatrix}
  0 &0&0\\
  \blacksquare &0 &0 \\
  \bullet & \bullet &\bullet
    \end{pmatrix}\,,\qquad
  \begin{pmatrix}
   0&0&0\\
   0&\blacksquare &0 \\
   \bullet& \bullet& \bullet
    \end{pmatrix}\,,
    \qquad
  \begin{pmatrix}
   0&0&0\\
   0&0 &\blacksquare  \\
   \bullet& \bullet&\bullet
    \end{pmatrix}\,.
\end{equation}
Here $\blacksquare$ stands for order $1$ element, $\bullet$ for (at
most) order $0.01$ element, while we neglect elements which are
$\lesssim 10^{-3}$. This particular hierarchy is enforced by a nontrivial
$a_\mu$ constraint that requires at least one large element in
the second row, while stringent upper bounds from LFV processes
exclude the possibility of having two elements of order 1.

We also identify the observables, which are most constraining for each
element in $Y$. We do this by registering the maximum increase in each
individual observable contribution to $\chi^2$ when a single $Y$
element is changed from its best-fit value. In the first row, all the
most stringent constraints actually also involve the $Y_{\mu q}$
elements: $K_L \to \mu^\pm e^\mp$ ($Y_{e d}$), $\mu\to e\gamma$ ($Y_{e
  s}$) and $B \to \pi \mu^\pm e^\mp$ ($Y_{e b}$).  In the near future,
we can expect some significant improvement at least for $Y_{e s}$ from
the MEG experiment~\cite{Dussoni:2009zz,Adam:2009ci}. The best-fit
regions around $\mathcal O(1)$ values for the second row elements are
mostly determined by the observed discrepancy in the $a_\mu$. Other
relevant observables, that constrain their values in the $\ll 1$
regions are: $\Delta m_K$ and $K_S \to \mu^+\mu^-$ ($Y_{\mu d}$, $Y_{\mu s}$), and
$B \to X_s \ell^+\ell^-$ ($Y_{\mu b}$). Unfortunately, due to the
theoretical uncertainties which dominate the precision of the first
two observables, a significant improvement in the foreseeable future
can only be expected for the constraint on $Y_{\mu b}$ from the Super
Flavor factories (SFFs)~\cite{O'Leary:2010af,Aushev:2010bq}. Finally,
the constraints on the third row of $Y$ are dominated by LFV tau and
$B$ decays: $\tau \to \mu \pi_0$ and  $\tau \to \mu \eta$ ($Y_{\tau
  d}$), $B \to K \tau^\pm \mu^\mp$ ($Y_{\tau s}$), and $B \to K
\tau^\pm \mu^\mp$ and $\tau \to \mu K_S$ ($Y_{\tau b}$)\,. Again SFFs
are expected to yield improved bounds on these LFV
observables. Finally we note the fact that bounds on most of the
elements of $Y$ are dominated by rare decays and the $a_\mu$ which all
exhibit a similar scaling dependence on the $\Delta$ parameters
($Y/m_\Delta$). This points towards an approximate linear scaling of
the fitted $Y$ element values with the $\Delta$ mass and allows for
simple reinterpretation of the derived limits at  $\Delta$ masses away
from the reference value $m_\Delta = 400$~GeV.

\subsection{Comment on tension between $\mathcal B({B \to \tau \nu})$ and $\sin 2\beta$}
\label{sec:Vub}

{ 

  We can redo the tree-level CKM fit, described in Sec.~\ref{sec:CKM}, 
  replacing the $|V_{ub}|$ value from Eq.~\eqref{eq:ckminputs} with a
  constraint coming from a world average of
  Belle and BaBar
  measurements~\cite{Ikado:2006un,:2008gx,Hara:2010dk,:2010rt} of $\mc{B} (B^+ \to \tau^+ \nu) = (1.68 \pm
  0.31)\E{-4}$~\cite{Lenz:2010gu}.
  The observable cannot be directly affected by $\Delta$ contributions and is given in the SM by
  \begin{equation}
  \mathcal B(B^+\to \tau^+ \nu_\tau ) = \frac{G_F^2 m_{B^+} m_\tau^2}{8\pi} \left(1-\frac{m_\tau^2}{m_{B^+}^2}\right)^2 |V_{ub}|^2 f_B^2 \tau_{B^+}\,.
  \end{equation}
  The main theoretical uncertainty due to the lattice QCD estimate of the relevant decay constant $f_B = 193 \pm 10\e{MeV}$~\cite{Laiho:2009eu} is at present subleading compared to the experimental error, but we nevertheless combine them in quadrature. The best-fitted values of the CKM parameters are then
  \begin{align}
    \label{eq:CKMparsTauNu}
    \lambda &= 0.22538(65)\,,\\
    A &=0.799(26)\,,\nonumber\\
    \rho &=0.162(90)\,,\nonumber\\
    \eta &=0.528(64)\,.\nonumber
  \end{align}
  The quality of the fit of the CKM from tree-level observables is exactly
  the same as in Sec.~\ref{sec:CKM}, however the central values of $\rho$
  and especially $\eta$ are significantly higher than before. This is
  expected since the tree level fit of $\rho$ and
  $\eta$ parameters is not over-constrained. We can also repeat the global fit of couplings $Y$,
  this time with a Gaussian chi-square term for the CKM matrix corresponding
  to Eq.~\eqref{eq:CKMparsTauNu} and the underlying correlation
  matrix.  The best fit point with $\chi^2 = 9.5$ relaxes the tension in
  the CKM by changing the $B_d-\bar B_d$ phase ($\sin 2\beta$) but at the price of not
  resolving the $a_\mu$ anomaly at all. Another, slightly shallower,
  minimum with $\chi^2 =10.6 $ achieves just the reverse -- $a_\mu$
  is perfectly satisfied while the tension in the CKM persists.
  
  A qualitative explanation goes as follows: a new phase in $B_d-\bar B_d$ mixing can be generated
  by either $Y_{e d} Y_{e b}^*$, $Y_{\mu d} Y_{\mu b}^*$ or $Y_{\tau d}
  Y_{\tau b}^*$.  Large $Y_{e d} Y_{e b}^*$ and $Y_{\mu d} Y_{\mu b}^*$ are ruled out by the strong $B \to \pi \ell^+ \ell^-$
  constraint. Thus for large enough $Y_{\tau d} Y_{\tau b}^*$ either
  (i) $Y_{\tau d}$ or (ii) $Y_{\tau b}$ is at least $\sim 0.1$. In turn we
  form uncomfortably
  large products of (i) $Y_{\tau d} Y_{\mu q}$ or (ii) $Y_{\tau b} Y_{\mu q}$ for $q=d,s,b$, where $Y_{\mu
    q}( \sim 1)$ is large to explain $a_\mu$.  First possibility
  $(q=d)$ is incompatible with (i) $\tau \to \pi^0 \mu$ or (ii) $B_s \to \tau \mu$ branching ratios,
  second one with (i) $\tau \to K^0 \mu$ or (ii) $B \to K \tau \mu$, and the last one with (i) $B_d \to \tau \mu$ or (ii) $\tau \to
  \mu \gamma$.

  Such worsening of the overall agreement of observables with the model can
  already be anticipated from Eqs.~\eqref{eq:Yranges} which
  clearly state that large contributions to $B_d-\bar B_d$ mixing are
  disfavoured. On the other hand, if one ignores the $a_\mu$
  constraint, then this model can sufficiently affect the phase of the $B_d-\bar B_d$
  mixing amplitude to be consistent with a large $\mc{B} (B \to \tau \nu)$.
}


\subsection{Comment on CPV in the $B_s$ system}
Next, we address the question whether
contributions of $\Delta$ can enhance the phase of the dispersive
amplitude in the $B_s-\bar B_s$ system, or even modify the
absorptive part by on-shell charged leptons in box
diagrams~(Fig.~\ref{fig:KKmixing}). According to~\cite{Dighe:2010nj},
which studied $\tau \tau$ absorbtive contributon to mixing amplitude,
one should have $Y_{\tau s} Y_{\tau b} \sim 0.1$ for $m_\Delta =
250\e{GeV}$. However, the set of observables we have included in
the analysis of $Y$ couplings (in particular $a_\mu$) forces the $\tau \tau$ and $\mu \mu$ states
to couple very weakly to $B_s$
\begin{subequations}
\begin{align}
  \label{eq:Bsbounds}
  |Y_{\mu s} Y^*_{\mu b}| &<0.0015\,(0.0021)\,,\\ 
  |Y_{\tau s} Y^*_{\tau b}| &< 0.9\E{-4}\,(4.1\E{-4})\,. \label{ytbytsbound}
\end{align}
\end{subequations}
Bounds are $1\,\sigma$\,($2\,\sigma$) C.L. (see also
Fig.~\ref{fig:Yx2Yx3})\,. While $|Y_{\mu s} Y^*_{\mu b}|$ is directly
constrained by the $B\to X_s \ell^+\ell^-$ rate, there is presently no
direct bound on the magnitude of $Y_{\tau s} Y^*_{\tau b}$, so the
constraint~\eqref{ytbytsbound} is directly linked to the explanation
of the $a_\mu$ anomaly. Using the above $2\,\sigma$ upper bounds in
Eq.~\eqref{eq:5} we find that the dispersive $\Delta$ amplitude with
tau (muon) in the box is five (four) orders of magnitude smaller than
the SM contribution.

The approximate scaling of the most important constraints with
$Y/m_\Delta$ provides a robust bound on the absorptive contributions
to the neutral meson mixing amplitudes, excluding any significant
modification of $\Delta \Gamma_s$, provided we require the resolution
of the $a_\mu$ anomaly. {We also note in passing that the smallness of
  new absorptive NP contributions is required in general by the
  measurements of the lifetime ratios of $B$ mesons, semileptonic
  branching fractions, and the average number of charm quarks in $B$
  decays (a recent discussion can be found in~\cite{Bobeth:2011st},
  see also ~\cite{Lenz:2010gu} and references therein).}  On the other
hand, the maximum allowed relative $\Delta$ contributions to the
dispersive parts scale quadratically with $m_{\Delta}$. In this way
dispersive $\Delta$ amplitudes comparable in size to SM contributions
in $B_s-\bar B_s$ mixing observables are only reached at masses well
above $1$~TeV, where the relevant $Y$ couplings are no longer
perturbative. Thus we find no possibility to simultaneously affect
$a_\mu$ and the $B_s$ system observables with $\Delta$ contributions.

\begin{figure}[tb]
  \centering
  \includegraphics[width=0.45\textwidth]{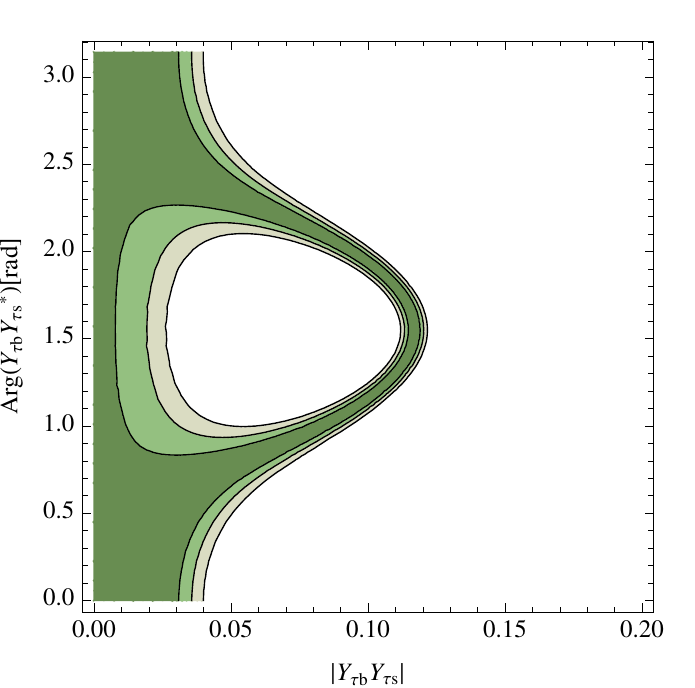}
  \caption{Correlation between the absolute value of $Y_{\tau s}
    Y_{\tau b}^*$ and its phase (${\rm Arg}(Y_{\tau s} Y_{\tau b}^*)$)
    in absence of the $a_{\mu}$ constraint. Dark green region is the
    1$\,\sigma$ contour, while the two lighter ones are 2 and
    3$\,\sigma$, respectively.}
  \label{fig:YtbYts}
\end{figure}

In absence of the $a_\mu$ constraint, the bounds on $Y_{\tau s}
Y_{\tau b}^*$ are significantly relaxed and are dominated by $\Delta
m_s$ and $\Delta m_s/\Delta m_d$ (see Fig.~\ref{fig:YtbYts}).
$|Y_{\tau s} Y_{\tau b}^*|$ values of order 0.1 are allowed, however
at the expense of fine-tuning the phase ${\rm Arg}(Y_{\tau s} Y_{\tau
  b}^*)$ in order to obtain the right destructive interference with
the SM contributions to the $B_{s,d}$ mass differences.

%
\section{GUT implications}
\label{sec:gutframework}
%

\subsection{Framework}
\label{framework}
The color triplet leptoquark $(\overline{\bm{3}}, \bm{1}, 4/3)$ emerges naturally in a theoretically
well-motivated class of grand unified models. We will first demonstrate this in a framework of the $SU(5)$ gauge group---the simplest group to encompass the SM gauge symmetry---and then proceed to discuss how and where it appears in the $SO(10)$ setup. 

\subsubsection{$SU(5)$ setup}

The matter of the SM is assigned to the $10$- and $5$-dimensional $SU(5)$ representations, i.e., $\bm{10}_i = (\bm{1},\bm{1},1)\oplus (\overline{\bm{3}},\bm{1},-2/3) \oplus (\bm{3},\bm{2},1/6)$ and $\bar{\bm 5}_i = (\bm{1},\bm{2},-1/2)\oplus(\overline{\bm{3}},\bm{1},1/3)$, where $i(=1,2,3)$ denotes generation index~\cite{Georgi:1974sy}. This assignment dictates that the charged fermion masses and the entries of the CKM matrix originate, at the tree-level, through the couplings of the matter fields to the $5$- and $45$-dimensional Higgs representations only~\cite{Slansky:1981yr}. It has actually been shown that the phenomenological considerations require presence of both~\cite{Georgi:1979df,Babu:1984vx,Giveon:1991zm,Dorsner:2006dj,Dorsner:2007fy,Perez:2007rm}. It turns out that the color triplet leptoquark is a part of the $45$-dimensional representation. Namely, the relevant SM decomposition reads $\bm{45}\equiv(\Delta_1, \Delta_2, \Delta_3, \Delta_4, \Delta_5,
\Delta_6, \Delta_7) = (\bm{8},\bm{2},1/2)\oplus
(\overline{\bm{6}},\bm{1}, -1/3) \oplus (\bm{3},\bm{3},-1/3)
\oplus (\overline{\bm{3}}, \bm{2}, -7/6) \oplus (\bm{3},\bm{1},
-1/3) \oplus (\overline{\bm{3}}, \bm{1}, 4/3) \oplus (\bm{1},
\bm{2}, 1/2)$. The color triplet thus appears in any $SU(5)$ 
framework that relies purely on the scalar representations for the charged fermion mass generation. 

Relevant contractions of the $45$- and $5$-dimensional Higgs representations, i.e., $\bm{45}$ and $\bm{5}$, with the matter fields, are $(Y_1)_{ij} \bm{10}_i \overline{\bm{5}}_j \bm{45}^{*}$, $(Y_2)_{ij} \bm{10}_i \bm{10}_j \bm{45}$, $(Y_3)_{ij} \bm{10}_i \overline{\bm{5}}_j \bm{5}^*$ and $(Y_4)_{ij} \bm{10}_i \bm{10}_j \bm{5}$, where $Y_a$, $a=1,2,3,4$, represent arbitrary Yukawa coupling matrices in flavor space. The charged fermion mass matrices at the unification scale accordingly read
\begin{eqnarray}
\label{massmatrices_D}
M_D &=&   - Y_1 v_{45}^* - \frac{1}{2} Y_3 v_5^* ,\\
\label{massmatrices_E}
M_E &=&  3 Y_1^T v_{45}^* - \frac{1}{2} Y_3^T v_5^* ,\\
\label{massmatrices_U}
M_U &=&  2 \sqrt{2}  (Y_2-Y^T_2) v_{45} - \sqrt{2} (Y_4+Y^T_4) v_5,
\end{eqnarray}
\noindent
where $\langle\bm{5}^5\rangle= v_5 / \sqrt{2}$ and $\langle\bm{45}^{15}_{1}\rangle= \langle\bm{45}^{2 5}_{2}\rangle=\langle\bm{45}^{3 5}_{3}\rangle = v_{45}/\sqrt{2}$ represent
appropriate vacuum expectation values.  Note that $\bm{5} \equiv \bm{5}^\alpha$, $\bm{45} \equiv \bm{45}^{\alpha \beta}_\gamma$ and $|v_5|^2/2+12 |v_{45}|^2=v^2$, where $\alpha, \beta, \gamma=1,..,5$ represent $SU(5)$ indices and $v(=246$\,GeV) stands for the electroweak vacuum expectation value (VEV). (The VEV result has been introduced for the first time in Ref.~\cite{Dorsner:2010cu} and corrects the normalization presented in Refs.~\cite{Dorsner:2007fy,Dorsner:2009cu}.) In $SU(5)$ there could be an additional contribution to $v$ from an $SU(2)$ triplet scalar~\cite{Dorsner:2005ii} but that contribution is supposed to be suppressed by a large symmetry breaking scale~\cite{Buras:1977yy} and we accordingly neglect it. We also assume that both $v_5$ and $v_{45}$ are real for simplicity. 

In order to have consistent notation we identify $\Delta_6$ with $\Delta$ in what follows. The lepton and baryon number violating Yukawa couplings of the triplet $\Delta$ to matter in the fermion mass eigenstate basis in the $SU(5)$ framework are already given in Eq.~\eqref{eq:lagr} if one makes the following identifications: $Y \equiv E_R^\dagger Y_1 D_R^*$ and $g \equiv  2 \sqrt{2} U_R^\dagger [Y_2-Y_2^T] U_R^*$. Here, $E_R$, $D_R$ and $U_R$ represent appropriate unitary transformations of the right-handed charged leptons, down-quarks and up-quarks. Our phenomenological study primarily relates to Yukawa couplings of $\Delta$ to the down-quark and charged lepton sectors. Clearly, these low-energy constraints on the leptoquark couplings to the matter could allow us to place constraints on the very Yukawa couplings and associated unitary transformations that show up in the charged fermion mass relations. These, on the other hand, might be pivotal in addressing the issue of matter stability~\cite{DeRujula:1980qc}.

Note that the antisymmetric nature of the color triplet couplings
to the up-quark sector in Eq.~\eqref{eq:lagr} is dictated by the group theory and is not affected by any change of basis. In other words, any unitary redefinition of fermion fields would preserve this property. We insist on this point for the following two reasons. Firstly, this is important since it is this unique feature of the $\Delta$ couplings to the up-quark sector that is
responsible for an absence of the leading contributions towards proton decay due to $\Delta$
exchange~\cite{Dorsner:2009cu}. Secondly, if, for some reason, $Y_2$ is a symmetric matrix, there would not be any coupling between $\Delta$ and the up sector. In other words, all $g_{ij}$ elements in Eq.~\eqref{eq:lagr} would be zero. If that was the case, $\Delta$ would not mediate proton decay. In fact that can happen, for example, if the scalar leptoquark $\Delta$ originates from an $SO(10)$ setup as we discuss next.

\subsubsection{$SO(10)$ setup}

Recall, one generation of the SM matter in the $SO(10)$ framework is embedded in a single $16$-dimensional representation. The allowed contractions of the matter fields to the Higgs sector, at the tree-level, are $(Y_{10})_{ij} \bm{16}_i \bm{16}_j \bm{10}$, $(Y_{120})_{ij} \bm{16}_i \bm{16}_j \bm{120}$ and $(Y_{126})_{ij} \bm{16}_i \bm{16}_j \bm{\overline{126}}$, where $\bm{10}$, $\bm{120}$ and $ \bm{\overline{126}}$ are the scalar representations that all contain states with the quantum numbers of the SM doublet~\cite{Slansky:1981yr}. Here, $Y_{10}(=Y_{10}^T)$, $Y_{120}(=-Y_{120}^T)$ and $Y_{126}(=Y_{126}^T)$ represent complex Yukawa coupling matrices. As it turns out, the $45$-dimensional representation of $SU(5)$ is found in both the $120$- and $126$-dimensional representations~\cite{Slansky:1981yr}. The former one couples antisymmetrically to matter, thus preserving the absence of the leading contributions towards proton decay due to $\Delta$ exchange~\cite{Dorsner:2009cu}. The latter one, on the other hand, couples symmetrically to matter. So, if $\Delta$ originates from the $126$-dimensional representation of $SO(10)$, it will not couple to the up-quark sector at all. Consequently, there will be no proton decay signatures related to $\Delta$ exchange in that case. Again, these properties are dictated by gauge symmetry and are preserved regardless of any redefinitions of the charged fermion fields. (Note that our findings on the absence of the up-quark sector couplings do not agree with the conclusions put forth in Ref.~\cite{Patel:2011eh} for the $SO(10)$ case and in Ref.~\cite{Perez:2008ry} for the $SU(5)$ case.) 

The relevant mass matrices for the down-quarks and charged leptons in the $SO(10)$ framework are
\begin{eqnarray}
\label{massmatrices_D_1}
M_D &=&   - Y_{126} v_{126}^* - \frac{1}{2} Y_{10} v_{10}^*+ Y_{120} (v_{120}'^{*}+v_{120}''^{*}),\\
\label{massmatrices_E_2}
M_E &=&  3 Y_{126} v_{126}^* - \frac{1}{2} Y_{10}^T v_{10}^* + Y_{120} (v_{120}'^{*}- 3 v_{120}''^{*}),
\end{eqnarray}
where $v_{10}$, $v_{126}$, $v_{120}'$ and $v_{120}''$ represent VEVs
of the doublet components of the appropriate scalar
representations. We will assume that the VEVs are real when needed for
simplicity. (See Ref.~\cite{Bajc:2005zf} for exact normalization with
respect to the SM VEV.) Clearly, the observed mismatch between the
charged lepton and down-quark masses requires a presence of either $\bm{120}$ or $\overline{\bm{126}}$, or both representations in the case without the $\bm{10}$. The color triplet hence must appear in any $SO(10)$ framework that relies purely on the scalar representations for the charged fermion mass generation.

We opt to start our analysis within a particular class of $SU(5)$ models having in mind that the same procedure can be carried over into an $SO(10)$ framework with appropriate modifications. In fact, towards the end of the next section we also address the $SO(10)$ setup viability in view of its compatibility with phenomenological constraints on the couplings of the light colored scalar to the matter fields. 

\subsection{Numerical Analysis}

Our goal is to consistently implement all available constraints on the color triplet couplings to the down-quarks and charged leptons in order to study implications for the charged fermion Yukawa sector within a particular class of grand unified models. These models rely solely on the scalar representations in order to generate charged fermion masses. 

We first single out a simple $SU(5)$ setup with the $5$-, $24$- and $45$-dimensional representations in the Higgs sector~\cite{Georgi:1979df,Babu:1984vx,Giveon:1991zm,Dorsner:2006dj,Dorsner:2007fy,Perez:2007rm} and one $24$-dimensional
fermionic representation~\cite{Perez:2007rm} to generate neutrino
masses via combination of type
I~\cite{Minkowski:1977sc,Yanagida:1979as,GellMann:1980vs,Glashow:1979nm,Mohapatra:1979ia}
and type III~\cite{Foot:1988aq,Ma:1998dn} seesaw mechanisms for definiteness. We resort to this model since it has been explicitly demonstrated that it predicts proton decay signatures that are very
close to the present experimental limits on the partial proton decay
lifetimes for the mass of $\Delta$ in the
range accessible in collider experiments~\cite{Dorsner:2009mq}. (The model is
a renormalizable version of the scenario first proposed
in~\cite{Bajc:2006ia} and further analyzed
in~\cite{Dorsner:2006fx,Bajc:2007zf}.) Moreover, it shares the same mass relations given in Eqs.~\eqref{massmatrices_D}, ~\eqref{massmatrices_E} and ~\eqref{massmatrices_U} with all other $SU(5)$ scenarios that rely on the use of the $5$- and $45$-dimensional scalar representations. 

We start with the following relations that are valid at the unification scale
\begin{eqnarray}
\label{1}
E_R^\dag D_L M^{\rm diag}_D &=&  (- \frac{1}{2} E_R^\dag Y_3 D_R^* v_5  - Y v_{45}) ,\\ \label{2}
M^{\rm diag}_E E_L^T D_R^*&=&  (- \frac{1}{2} E_R^\dag Y_3 D_R^* v_5 + 3 Y v_{45}),
\end{eqnarray}
where $M^{\rm diag}_D$ and $M^{\rm diag}_E$ are diagonal mass matrices for down quarks and charged leptons, respectively. Our convention is such that $M_D=D_L M^{\rm diag}_D D_R^T$ and $M_E=E_L M^{\rm diag}_E E_R^T$, where $D_L$ and $E_L$ represent appropriate unitary transformations of the left-handed down quarks and charged leptons. Note that our phenomenological considerations yield constrains on the form of $Y$ that are valid at low energies only. It is thus essential to propagate constraints on $(Y)_{ij} $, $i,j=1,2,3$, as well as the entries of $M^{\rm diag}_D$ and $M^{\rm diag}_E$ to the GUT scale to extract accurate information on $v_{45}$ and unitary matrices $E_R^\dag D_L$ and $E_L^T D_R^*$.

Again, the phenomenological bounds we derive constrain the matrix $Y$ appearing on the right-hand side of a relation 
\begin{equation}
\label{main}
E_R^\dag D_L M^{\rm diag}_D-M^{\rm diag}_E E_L^T D_R^*=-4 Y v_{45}.
\end{equation} 
What is not known are the overall scale of the right-hand side set by $v_{45}$ and the unitary transformations given by $E_R^\dag D_L$ and $E_L^T D_R^*$. In order to perform numerical analysis and implement inferred bounds we first parametrize $E_R^\dag D_L$ and $E_L^T D_R^*$ using a generic form 
\begin{equation}
\label{eq:PMNS}
U=\begin{pmatrix}
e^{i\alpha_1} & 0 & 0\\
0 & e^{i\alpha_2} & 0\\
0 & 0 & e^{i\alpha_3}
\end{pmatrix} \begin{pmatrix}
c_{12}c_{13} & s_{12}c_{13} & s_{13}e^{-i\alpha_4} \\
-s_{12}c_{23}-c_{12}s_{23}s_{13}e^{i\alpha_4} &
c_{12}c_{23}-s_{12}s_{23}s_{13}e^{i\alpha_4} &  s_{23}c_{13}\\
s_{12}s_{23}-c_{12}c_{23}s_{13}e^{i\alpha_4} &
-c_{12}s_{23}-s_{12}c_{23}s_{13}e^{i\alpha_4} & c_{23}c_{13}
\end{pmatrix}
\begin{pmatrix}
e^{i\alpha_5} & 0 & 0\\
0 & e^{i\alpha_6} & 0\\
0 & 0 & 1
\end{pmatrix},
\end{equation}
where $s_{ab}\equiv\sin\theta_{ab}$, $c_{ab}\equiv\cos\theta_{ab}$, and $\alpha_i$, $i=1,..,6$, are
phases. We then randomly generate the total of nineteen parameters and
check whether the left-hand side of Eq.~\eqref{main} satisfies all
phenomenological constraints. (We also vary the four parameters of the
CKM matrix as well as $\eta_1$, $\eta_2$ and $\eta_3$ --- QCD parameters entering $K-\bar{K}$ mixing---in order to have consistent constraints on the $Y$ entries as described in Sec.~\ref{sec:pheno}.) This process is repeated until the available parameter space is thought to be exhausted. The down-quark and charged lepton masses at the GUT scale are considered as input and the relevant values we generate and use within this particular framework are given in Table~\ref{tab:input}. The GUT scale is taken to be $M_{\rm GUT}=10^{16}$\,GeV and we only propagate and use the central values for the down-quark and charged lepton masses.
\begin{table}[tb]
\begin{tabular}{ll}
\hline
running mass at $M_Z$ $\qquad \qquad \qquad$
& running mass at $M_{\rm GUT}$\qquad \qquad  \\ \hline
$m_b(M_Z) =  2.89 \pm 0.11$\,GeV & $m_b(M_{\rm GUT}) =  0.782$\,GeV\\
$m_s(M_Z) =   56 \pm 16 $\,MeV & $m_s(M_{\rm GUT}) =  19 $\,MeV\\
$m_d(M_Z) =  3.0 \pm 1.2$\,MeV & $m_d(M_{\rm GUT}) =  1.1$\,MeV\\
$m_{\tau}(M_Z) =
1746.45^{+0.29}_{-0.26}$\,MeV & $m_{\tau}(M_{\rm GUT}) =
1561.4$\,MeV\\
$m_{\mu}(M_Z) = 102.72899(44)$\,MeV & $m_{\mu}(M_{\rm GUT}) = 91.84$\,MeV\\
$m_{e}(M_Z)   = 0.4866613(36)$\,MeV & $m_{e}(M_{\rm GUT})   = 0.4350$\,MeV\\
\hline
\end{tabular}
\caption{\label{tab:input}
Input parameters for the relevant fermion masses at the $M_Z$ scale and the corresponding values at the GUT scale ($M_{\rm GUT}=10^{16}$\,GeV) in a non-supersymmetric framework.}
\end{table}

Note that the need to accommodate experimental results on $a_\mu$ basically sets the scale for the $Y$ entries. To be precise, it requires that $\sum_{i=1,2,3}|Y_{2i}|^2$ satisfies Eq.~\eqref{eq:amu}. This in turn should fix the value or range of allowed values of $v_{45}$ since the scale of the left-hand side of Eq.~\eqref{main} is set by the known fermion masses. One can then use this information to determine $v_{5}$ via $|v_5|^2/2+12 |v_{45}|^2=v^2$. To be conservative we not only vary $\sum_{i=1,2,3}|Y_{2i}|^2$ within the 1\,$\sigma$ and 2\,$\sigma$ ranges but accommodate for the effect of the RGE running  of our constraints from the low scale to the grand unified scale. We take that effect to be within the bounds set by the following scaling factors: $1.1$--$3.7$. These scaling factors correspond to the maximal changes in the charged lepton and down quark masses as they are propagated from low scale to the GUT scale. Again, we take the GUT scale to be $M_{\rm GUT}=10^{16}$\,GeV for simplicity. (The exact dependence of the GUT scale on the scalar particle mass spectrum within this particular $SU(5)$ model is known and has been worked out in detail in Ref.~\cite{Dorsner:2009mq}. The change in the GUT scale or, correspondingly, the scalar particle mass spectrum also affects propagation of fermion masses but that effect is rather small for the scenario when $\Delta$ is light as the GUT scale is then limited within a very narrow range~\cite{Dorsner:2009mq}.) 

\begin{figure}[h]
\includegraphics[width=0.80\textwidth]{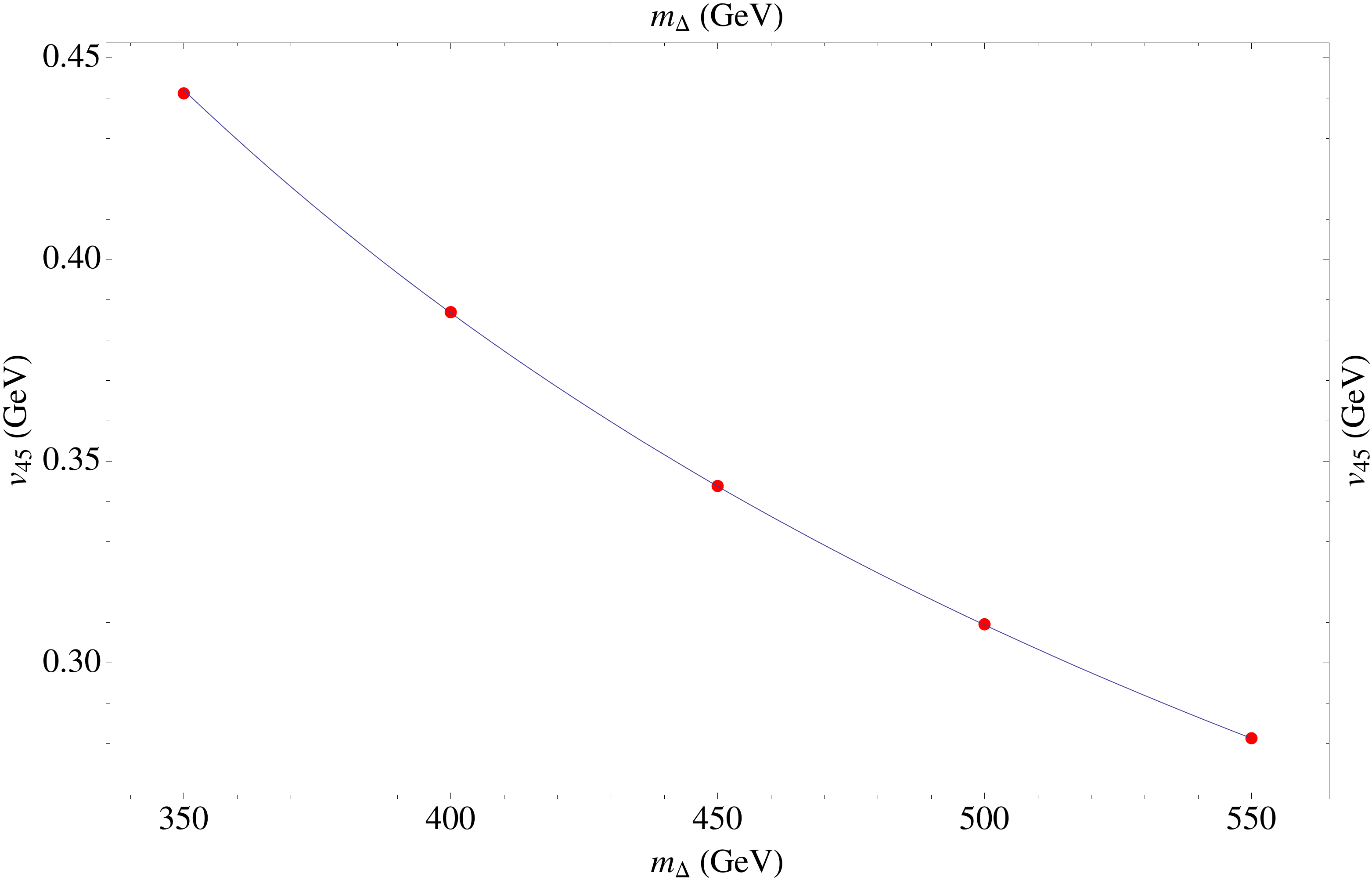}
\caption{Upper bound on $v_{45}$ as a function of $m_{\Delta}$. Data are generated for a discrete set of $m_{\Delta}$ values that are shown as dots. The curve is an interpolation that carries an $m_{\Delta}^{-1}$ dependence.}
\label{fig1}
\end{figure}

The upper limit on $v_{45}$ which we obtain by randomly choosing the entries of $E_R^\dag D_L$ and $E_L^T D_R^*$ is shown in Fig.~\ref{fig1}. Clearly, the bound should drop as $m_{\Delta}$ grows since $v_{45}$ needs to compensate the growth of the appropriate values of $Y$ that satisfy the $a_{\mu}$ constraint of Eq.~\eqref{eq:amu}.  For practical purposes, we generate this conservative limit when only Eq.~\eqref{eq:amu} is satisfied for a finite set of fixed values of $m_\Delta$. These correspond to dots in Fig.~\ref{fig1}. In our numerical study we limit the $m_{\Delta}$ range due to the existence of both the lower and upper bounds on its value. The lower experimental bound on $m_{\Delta}$ comes from direct experimental searches. The most stringent one originates from dedicated searches for pair production of leptoquarks in $p\,p$ collisions at LHC and it reads $m_{\Delta} > 384$\,GeV~\cite{Khachatryan:2010mp} ($m_{\Delta} > 422$\,GeV~\cite{Aad:2011uv}) for the so-called first-generation (second-generation) leptoquarks assuming these decay exclusively to an electron (muon) and a hadronic jet. While these bounds are not necessarily applicable to our framework, since $\Delta$ can also decay to a top quark and a hadronic jet, we have verified that the corresponding branching ratio is always below 30\% in the region of parameter space where $\Delta$ resolves both the $t\bar{t}$ FBA and the $a_\mu$ puzzles. The upper bound on $m_{\Delta}$, on the other hand, originates from perturbativity constraints on entries of $Y$ that should not exceed $\sqrt{4 \pi}$. We find that bound to be $m_{\Delta} \lesssim 560$\,GeV.

After an extensive numerical study we fail to generate a single satisfactory solution to all the constraints using Eq.~\eqref{main} as a starting point. We trace the difficulty of finding a viable numerical solution to the facts that (i) the down-quark and charged lepton sectors do not exhibit  a strong mass hierarchy that is present in the up-quark sector and (ii) the misalignment between the masses of the down-quarks and charged leptons that belong to the same generation is sufficiently large to prevent necessary cancellations. For example, a generic form of the left-hand side in Eq.~\eqref{main} can be represented as follows
\begin{equation}
  \begin{pmatrix}
   0&0&0\\
   0&0 &0  \\
   \blacksquare& \blacksquare & \blacksquare
    \end{pmatrix}\,+
      \begin{pmatrix}
   0&0&\bullet\\
   0&0 &\bullet  \\
   0& 0 & \bullet
    \end{pmatrix}\,.
\end{equation}
Here $\blacksquare$ ($\bullet$) stands for an order $m_\tau$ ($m_b$) element. Clearly, the only potentially viable scenario for this form to describe matrix $Y$, pictorially given in Eq.~\eqref{eq:Yhierarchy}, would be the one where the 23 element dominates. The 31 and 32 elements should accordingly be suppressed by effectively setting the angles $\theta_{13}$ and $\theta_{23}$ from $E_L^T D_R^*$ to zero. This, however, leaves the 33 element on the left-hand side of Eq.~\eqref{main} to be proportional to $m_\tau-m_b (E_R^\dag D_L)_{33}$. As $m_\tau(M_{\rm GUT}) \sim 2 m_b(M_{\rm GUT})$ in the scenario at hand and $|(E_R^\dag D_L)_{33}| \leq 1$, the absolute value of the 33 element turns out to always be greater than the absolute value of the 23 element, in contrast to what is needed. One could try to see if there is a possibility to have a satisfactory numerical solution within the supersymmetric framework where, for example, the mismatch between $b$ and $\tau$ varies a lot with the change in the $\tan \beta$ parameter. This scenario, although it does help in suppressing the 33 element, also fails due to the difficulty to accommodate small enough elements in the 1-2 block of the left-hand side of Eq.~\eqref{main}. Namely, once the freedom to set the 13 and 33 elements to be small by tuning the angles in $E_R^\dag D_L$ is used there is not enough parameters left over to tune the 1-2 block to the desired form. For example, since $m_{e}(M_{\rm GUT})/m_{b}(M_{\rm GUT}) \sim m_{d}(M_{\rm GUT})/m_{b}(M_{\rm GUT}) \sim 10^{-4}$, the 11 element is always bigger than the required limit of $10^{-6}$. In short, the $SU(5)$ scenarios with a light triplet scalar that rely on the use of the $5$- and $45$-dimensional scalar representations to generate charged fermion masses at the tree level fail to accommodate the Yukawa structure needed to explain the $a_\mu$ puzzle while satisfying all other phenomenological constraints. 

Let us now discuss implications of our findings with respect to their compatibility with the most commonly encountered $SO(10)$ scenarios. Recall, the only representations of $SO(10)$ that could, at the tree-level, yield contributions to the charged fermion masses are the $\bm{10}$, $\bm{120}$ and $\overline{\bm{126}}$. And, as we have pointed out in Sec.~\ref{framework}, $\Delta$ can originate from either $120$- or $126$-dimensional representation of $SO(10)$.

If $\Delta$ is part of the $126$-dimensional Higgs it would not couple to the up-quark sector since the relevant couplings to matter are symmetric whereas $\Delta$ needs to couple in an antisymmetric manner to the up-quarks. If, in addition to the $\overline{\bm{126}}$, one uses a $10$-dimensional scalar representation to generate the charged fermion masses the corresponding mass matrices will all be symmetric. This, on the other hand, changes the transformations that generate mass eigenstate basis from bi-unitary into congruent form. This significantly reduces the number of free parameters yielding the following mass relation
\begin{equation}
\label{main1}
U M^{\rm diag}_D-M^{\rm diag}_E U^*=-4 Y v_{126},
\end{equation} 
where $U=E_R^\dag D_R$, $E_R=E_L$ and $D_R=D_L$. This relation also corresponds to the $SU(5)$ scenario when all Yukawa matrices in the down-quark and charged lepton sectors are symmetric. Obviously, this case is much more restrictive since we have only one unitary matrix $U$ to vary. It is thus clear that this scenario cannot be viable if we implement all the constraints on the form of $Y$. Hence, the case when Yukawa couplings in the charged lepton and down-quark sectors are symmetric, including the case with the $10$- and $126$-dimensional scalar representations in $SO(10)$, is not compatible with possibility to have light $\Delta$ as an explanation for observed anomalies. 

The scenario with the $10$- and $120$-dimensional representations in the Higgs sector is also not realistic. In fact, that scenario resembles  the $SU(5)$ scenario that proved to be inadequate to accommodate the form of $Y$ matrix. Moreover, the $10$- and $120$-dimensional representation scenario cannot explain observed fermion masses as was demonstrated in the low-scale supersymmetric case~\cite{Lavoura:2006dv}. In fact, even the $\overline{\bm{126}}$ and $\bm{10}$ of Higgs scenario would require complex $\bm{10}$ just to meet the charged fermion mass constraints~\cite{Bajc:2005zf} in the non-supersymmetric case. This finally leaves, as the only viable possibility, the most general scenario with the $10$-, $120$- and $126$-dimensional representations as the one that could accommodate constraints generated by the $\Delta$ phenomenology in the $SO(10)$ framework. The relevant relation, in that scenario, reads
\begin{equation}
\label{main2}
E_R^\dag D_L M^{\rm diag}_D-M^{\rm diag}_E E_L^T D_R^*=-4 E_R^\dagger Y_{126} D_R^* v_{126}+4 E_R^\dagger Y_{120} D_R^* v''_{120}.
\end{equation} 
Clearly, $E_R^\dagger Y_{126} D_R^*$ ($E_R^\dagger Y_{120} D_R^*$) would be proportional to $Y$ for $\Delta$ originating from $\overline{\bm{126}}$ ($\bm{120}$). In both cases there are more than enough parameters to accommodate required form of $Y$. Note, however, that our conservative estimate for the upper bound on $v_{45}$ as shown in Fig.~\ref{fig1} should still be applicable on either $v_{126}$ or $v''_{120}$. 
For example, if we identify $E_R^\dagger Y_{126} D_R^*$ ($E_R^\dagger Y_{120} D_R^*$) with $Y$ it is clear that the left-hand side of Eq.~\eqref{main2} cannot be dominated by the term proportional to $v_{120}''$ ($v_{126}$). If the opposite was true, we would obtain $E_R^\dagger Y_{126} D_R^* \sim E_R^\dagger Y_{120} D_R^*$ which certainly cannot hold as $Y_{126}$ is symmetric and $Y_{120}$ is antisymmetric. To conclude, the only viable candidate that can accommodate $Y$ is the $SO(10)$ scenario with the $\bm{10}$, $\bm{120}$ and $\overline{\bm{126}}$.

%
\section{Conclusions}
\label{sec:conclusions}
%

We have investigated the role of a colored weak singlet scalar possibly addressing the 
$t\bar t$ FBA puzzle in flavor changing processes and precision observables of down-quarks and
charged leptons. The magnitude of the predicted effects is governed by the mass of
the scalar (which we normalize to $400$\,GeV as preferred by the $t\bar t$ phenomenology), and  (generic)
complex matrix $Y$ acting in quark and lepton flavor-space. $Y$ is the central object of this analysis.

Virtual contributions of the considered scalar affect many observables and
in order to obtain insight into the $Y$ structure we have analyzed a plethora
of rare quark and lepton processes, some of them well measured,
others bounded from above.  In particular we have considered FCNC
and CP violating observables in $K$ and $B_{d,s}$ meson systems, (lepton flavor
violating) dileptonic decays of neutral mesons, $\mu-e$ conversion
in nuclei, anomalous magnetic moments of charged leptons, and
lepton flavor violating decays of the muon and $\tau$ lepton. For completeness, we have also
considered effects in the $Z\to b \bar b$ decay width. We have properly
accounted for SM contributions to the relevant observables where needed.

Then we have performed a global $\chi^2$ fit of the $Y$ matrix
elements and found an excellent agreement with all the considered
constraints. We have confirmed that the couplings to electrons are
strongly suppressed. The most salient finding is the explanation of
the anomalous magnetic moment of muon, which requires the muon
coupling to a single generation down-quark to be of order one.
Combined with LFV $B$ and $\tau$ decay constraints, this leads to
strong limits on the tau lepton couplings to down quarks which in turn
exclude the possibility~\cite{Dighe:2010nj} to simultaneously explain
the measured large CP-violating mixing phase in the $B_s$ sector or a
large enhancement of absorptive mixing amplitude $\Gamma_{12s}$ in
this model. Even in absence of the $a_\mu$ constraint, the
$B_s-\bar B_s$ mass difference measurements alone constrain the
relevant leptoquark couplings and a large new absorptive contribution
in $B_s-\bar B_s$ mixing cannot be generated. {Using a value of
  $|V_{ub}|$ preferred by the measured branching fraction of $B \to \tau \nu$ we find that this
  model can modify the $B_d-\bar B_d$ mixing amplitude sufficiently to
  remove the tension between the two observables.  However, in this
  case the anomalous magnetic moment of the muon cannot be explained.}

We have systematically implemented all the phenomenological constraints in a class of $SU(5)$ models where all the fermion masses are generated at the tree-level to find out that the explanation of the $a_\mu$ anomaly requires the vacuum expectation value of the $45$-dimensional representation to be of the order of $10^{-1}$\,GeV. This result implies that the up-quark couplings, in this setup, are symmetric in nature. This in turn makes predictions for certain partial proton decay lifetimes very accurate.  We have also shown that the symmetric scenario for the Yukawa couplings in the down-quark and charged lepton case is not compatible with the constraints due to the presence of light $\Delta$ and discussed implications for the $SO(10)$ type of unification. The simplest of possible realizations of both $SO(10)$ and $SU(5)$ with the symmetric Yukawa sector, that could accommodate observed fermion masses, are shown not to be viable unless $\Delta$ is heavy enough not to play any role in low-energy phenomenology. 

We conclude by noting, that the couplings of the leptoquark in
question to the matter fields, in the physical basis, are always
dominated by just one of the entries of the second row of matrix
$Y$. That entry is at least two orders of magnitude larger than any
other entry. This property puts this particular leptoquark effectively
in the so-called second generation category. Moreover, as it does not
couple to neutrinos, the bound extracted from the recent LHC data  for
the second-generation leptoquarks~\cite{Aad:2011uv} is truly
applicable in this case and reads $m_{\Delta} \gtrsim 380$\,GeV,
accounting for the reduced $\Delta \to \mu j$ branching ratio of order
$\mathcal B\gtrsim 0.7$ due to the presence of the $\Delta \to t j$
decay channel~\cite{Dorsner:2009mq}. This and the upper bound on its
mass --- $m_{\Delta} < 560$\,GeV --- that  originates from simple perturbativity arguments thus place it in a very narrow window of discovery.

\begin{acknowledgments}
  We thank Ulrich Haisch for pointing out several important $B$ decay
  modes missing in the first version of the analysis and for useful
  discussions regarding the $B_s$ mixing phenomenology. We also
  acknowledge useful discussions with Jure Zupan. N.K. thanks Fran\c
  cois Le Diberder for invaluable advice on the fit part. I.D. thanks
  the Institut ``Jo\v zef Stefan" where part of this work was
  completed for their hospitality.  This work is supported in part by
  the Slovenian Research Agency.
\end{acknowledgments}

\bibliography{./refs}

\begin{thebibliography}{96}
\expandafter\ifx\csname natexlab\endcsname\relax\def\natexlab#1{#1}\fi
\expandafter\ifx\csname bibnamefont\endcsname\relax
  \def\bibnamefont#1{#1}\fi
\expandafter\ifx\csname bibfnamefont\endcsname\relax
  \def\bibfnamefont#1{#1}\fi
\expandafter\ifx\csname citenamefont\endcsname\relax
  \def\citenamefont#1{#1}\fi
\expandafter\ifx\csname url\endcsname\relax
  \def\url#1{\texttt{#1}}\fi
\expandafter\ifx\csname urlprefix\endcsname\relax\def\urlprefix{URL }\fi
\providecommand{\bibinfo}[2]{#2}
\providecommand{\eprint}[2][]{\url{#2}}

\bibitem[{\citenamefont{Kamenik et~al.}(2011)\citenamefont{Kamenik, Shu, and
  Zupan}}]{topReview}
\bibinfo{author}{\bibfnamefont{J.~F.} \bibnamefont{Kamenik}},
  \bibinfo{author}{\bibfnamefont{J.}~\bibnamefont{Shu}}, \bibnamefont{and}
  \bibinfo{author}{\bibfnamefont{J.}~\bibnamefont{Zupan}}
  (\bibinfo{year}{2011}), \eprint{1107.5257}.

\bibitem[{\citenamefont{Dorsner
  et~al.}(2010{\natexlab{a}})\citenamefont{Dorsner, Fajfer, Kamenik, and
  Kosnik}}]{Dorsner:2009mq}
\bibinfo{author}{\bibfnamefont{I.}~\bibnamefont{Dorsner}},
  \bibinfo{author}{\bibfnamefont{S.}~\bibnamefont{Fajfer}},
  \bibinfo{author}{\bibfnamefont{J.~F.} \bibnamefont{Kamenik}},
  \bibnamefont{and} \bibinfo{author}{\bibfnamefont{N.}~\bibnamefont{Kosnik}},
  \bibinfo{journal}{Phys. Rev.} \textbf{\bibinfo{volume}{D81}},
  \bibinfo{pages}{055009} (\bibinfo{year}{2010}{\natexlab{a}}),
  \eprint{0912.0972}.

\bibitem[{\citenamefont{Gresham
  et~al.}(2011{\natexlab{a}})\citenamefont{Gresham, Kim, and
  Zurek}}]{Gresham:2011pa}
\bibinfo{author}{\bibfnamefont{M.~I.} \bibnamefont{Gresham}},
  \bibinfo{author}{\bibfnamefont{I.-W.} \bibnamefont{Kim}}, \bibnamefont{and}
  \bibinfo{author}{\bibfnamefont{K.~M.} \bibnamefont{Zurek}}
  (\bibinfo{year}{2011}{\natexlab{a}}), \eprint{1103.3501}.

\bibitem[{\citenamefont{Blum et~al.}(2011)\citenamefont{Blum, Hochberg, and
  Nir}}]{Blum:2011fa}
\bibinfo{author}{\bibfnamefont{K.}~\bibnamefont{Blum}},
  \bibinfo{author}{\bibfnamefont{Y.}~\bibnamefont{Hochberg}}, \bibnamefont{and}
  \bibinfo{author}{\bibfnamefont{Y.}~\bibnamefont{Nir}} (\bibinfo{year}{2011}),
  \eprint{1107.4350}.

\bibitem[{\citenamefont{Gresham
  et~al.}(2011{\natexlab{b}})\citenamefont{Gresham, Kim, and
  Zurek}}]{Gresham:2011fx}
\bibinfo{author}{\bibfnamefont{M.~I.} \bibnamefont{Gresham}},
  \bibinfo{author}{\bibfnamefont{I.-W.} \bibnamefont{Kim}}, \bibnamefont{and}
  \bibinfo{author}{\bibfnamefont{K.~M.} \bibnamefont{Zurek}}
  (\bibinfo{year}{2011}{\natexlab{b}}), \eprint{1107.4364}.

\bibitem[{\citenamefont{Dorsner
  et~al.}(2010{\natexlab{b}})\citenamefont{Dorsner, Fajfer, Kamenik, and
  Kosnik}}]{Dorsner:2010cu}
\bibinfo{author}{\bibfnamefont{I.}~\bibnamefont{Dorsner}},
  \bibinfo{author}{\bibfnamefont{S.}~\bibnamefont{Fajfer}},
  \bibinfo{author}{\bibfnamefont{J.~F.} \bibnamefont{Kamenik}},
  \bibnamefont{and} \bibinfo{author}{\bibfnamefont{N.}~\bibnamefont{Kosnik}},
  \bibinfo{journal}{Phys. Rev.} \textbf{\bibinfo{volume}{D82}},
  \bibinfo{pages}{094015} (\bibinfo{year}{2010}{\natexlab{b}}),
  \eprint{1007.2604}.

\bibitem[{\citenamefont{Georgi}(1975)}]{Georgi:1975qb}
\bibinfo{author}{\bibfnamefont{H.}~\bibnamefont{Georgi}}
  (\bibinfo{year}{1975}), \bibinfo{note}{in the Proceedings of Theories and
  Experiments in High-Energy Physics, Center for Theoretical Physics: Univ. of
  Miami.}

\bibitem[{\citenamefont{Fritzsch and Minkowski}(1975)}]{Fritzsch:1974nn}
\bibinfo{author}{\bibfnamefont{H.}~\bibnamefont{Fritzsch}} \bibnamefont{and}
  \bibinfo{author}{\bibfnamefont{P.}~\bibnamefont{Minkowski}},
  \bibinfo{journal}{Annals Phys.} \textbf{\bibinfo{volume}{93}},
  \bibinfo{pages}{193} (\bibinfo{year}{1975}).

\bibitem[{\citenamefont{Patel and Sharma}(2011)}]{Patel:2011eh}
\bibinfo{author}{\bibfnamefont{K.~M.} \bibnamefont{Patel}} \bibnamefont{and}
  \bibinfo{author}{\bibfnamefont{P.}~\bibnamefont{Sharma}},
  \bibinfo{journal}{JHEP} \textbf{\bibinfo{volume}{1104}}, \bibinfo{pages}{085}
  (\bibinfo{year}{2011}), \eprint{1102.4736}.

\bibitem[{\citenamefont{Bennett et~al.}(2004)}]{Bennett:2004pv}
\bibinfo{author}{\bibfnamefont{G.~W.} \bibnamefont{Bennett}}
  \bibnamefont{et~al.} (\bibinfo{collaboration}{Muon g-2}),
  \bibinfo{journal}{Phys. Rev. Lett.} \textbf{\bibinfo{volume}{92}},
  \bibinfo{pages}{161802} (\bibinfo{year}{2004}), \eprint{hep-ex/0401008}.

\bibitem[{\citenamefont{Jegerlehner}(2007)}]{Jegerlehner:2007xe}
\bibinfo{author}{\bibfnamefont{F.}~\bibnamefont{Jegerlehner}},
  \bibinfo{journal}{Acta Phys. Polon.} \textbf{\bibinfo{volume}{B38}},
  \bibinfo{pages}{3021} (\bibinfo{year}{2007}), \eprint{hep-ph/0703125}.

\bibitem[{\citenamefont{Aaltonen et~al.}(2008)}]{Aaltonen:2007he}
\bibinfo{author}{\bibfnamefont{T.}~\bibnamefont{Aaltonen}} \bibnamefont{et~al.}
  (\bibinfo{collaboration}{CDF Collaboration}),
  \bibinfo{journal}{Phys.Rev.Lett.} \textbf{\bibinfo{volume}{100}},
  \bibinfo{pages}{161802} (\bibinfo{year}{2008}), \eprint{0712.2397}.

\bibitem[{\citenamefont{Abazov et~al.}(2008)}]{:2008fj}
\bibinfo{author}{\bibfnamefont{V.~M.} \bibnamefont{Abazov}}
  \bibnamefont{et~al.} (\bibinfo{collaboration}{D0}), \bibinfo{journal}{Phys.
  Rev. Lett.} \textbf{\bibinfo{volume}{101}}, \bibinfo{pages}{241801}
  (\bibinfo{year}{2008}), \eprint{0802.2255}.

\bibitem[{\citenamefont{{}}(2010{\natexlab{a}})}]{D0betasUpdate}
\bibinfo{author}{\bibnamefont{{}}} (\bibinfo{collaboration}{D0 Collaboration})
  (\bibinfo{year}{2010}{\natexlab{a}}), \eprint{{Note 6093-CONF}}.

\bibitem[{\citenamefont{{}}(2010{\natexlab{b}})}]{CDFbetasUpdate}
\bibinfo{author}{\bibnamefont{{}}} (\bibinfo{collaboration}{CDF Collaboration})
  (\bibinfo{year}{2010}{\natexlab{b}}), \eprint{{Public Note 10206}}.

\bibitem[{\citenamefont{Abazov et~al.}(2010)}]{Abazov:2010hv}
\bibinfo{author}{\bibfnamefont{V.~M.} \bibnamefont{Abazov}}
  \bibnamefont{et~al.} (\bibinfo{collaboration}{D0 Collaboration}),
  \bibinfo{journal}{Phys.Rev.} \textbf{\bibinfo{volume}{D82}},
  \bibinfo{pages}{032001} (\bibinfo{year}{2010}), \eprint{1005.2757}.

\bibitem[{\citenamefont{Abazov et~al.}(2011)}]{Abazov:2011yk}
\bibinfo{author}{\bibfnamefont{V.~M.} \bibnamefont{Abazov}}
  \bibnamefont{et~al.} (\bibinfo{collaboration}{D0 Collaboration})
  (\bibinfo{year}{2011}), \eprint{1106.6308}.

\bibitem[{\citenamefont{Ligeti et~al.}(2010)\citenamefont{Ligeti, Papucci,
  Perez, and Zupan}}]{Ligeti:2010ia}
\bibinfo{author}{\bibfnamefont{Z.}~\bibnamefont{Ligeti}},
  \bibinfo{author}{\bibfnamefont{M.}~\bibnamefont{Papucci}},
  \bibinfo{author}{\bibfnamefont{G.}~\bibnamefont{Perez}}, \bibnamefont{and}
  \bibinfo{author}{\bibfnamefont{J.}~\bibnamefont{Zupan}},
  \bibinfo{journal}{Phys. Rev. Lett.} \textbf{\bibinfo{volume}{105}},
  \bibinfo{pages}{131601} (\bibinfo{year}{2010}), \eprint{1006.0432}.

\bibitem[{\citenamefont{Lenz et~al.}(2011)}]{Lenz:2010gu}
\bibinfo{author}{\bibfnamefont{A.}~\bibnamefont{Lenz}} \bibnamefont{et~al.},
  \bibinfo{journal}{Phys. Rev.} \textbf{\bibinfo{volume}{D83}},
  \bibinfo{pages}{036004} (\bibinfo{year}{2011}), \eprint{1008.1593}.

\bibitem[{\citenamefont{Grossman}(1996)}]{Grossman:1996era}
\bibinfo{author}{\bibfnamefont{Y.}~\bibnamefont{Grossman}},
  \bibinfo{journal}{Phys. Lett.} \textbf{\bibinfo{volume}{B380}},
  \bibinfo{pages}{99} (\bibinfo{year}{1996}), \eprint{hep-ph/9603244}.

\bibitem[{\citenamefont{Saha et~al.}(2010)\citenamefont{Saha, Misra, and
  Kundu}}]{Saha:2010vw}
\bibinfo{author}{\bibfnamefont{J.~P.} \bibnamefont{Saha}},
  \bibinfo{author}{\bibfnamefont{B.}~\bibnamefont{Misra}}, \bibnamefont{and}
  \bibinfo{author}{\bibfnamefont{A.}~\bibnamefont{Kundu}},
  \bibinfo{journal}{Phys.Rev.} \textbf{\bibinfo{volume}{D81}},
  \bibinfo{pages}{095011} (\bibinfo{year}{2010}), \eprint{1003.1384}.

\bibitem[{\citenamefont{Bona et~al.}(2010)}]{UTfit}
\bibinfo{author}{\bibfnamefont{M.}~\bibnamefont{Bona}} \bibnamefont{et~al.}
  (\bibinfo{collaboration}{UTfit}), \bibinfo{journal}{JHEP}
  \textbf{\bibinfo{volume}{03}}, \bibinfo{pages}{080} (\bibinfo{year}{2010}).

\bibitem[{\citenamefont{Nakamura et~al.}(2010)}]{Nakamura:2010zzi}
\bibinfo{author}{\bibfnamefont{K.}~\bibnamefont{Nakamura}} \bibnamefont{et~al.}
  (\bibinfo{collaboration}{Particle Data Group}), \bibinfo{journal}{J. Phys.}
  \textbf{\bibinfo{volume}{G37}}, \bibinfo{pages}{075021}
  (\bibinfo{year}{2010}).

\bibitem[{\citenamefont{{}}(2010{\natexlab{c}})}]{:1900yx}
\bibinfo{author}{\bibnamefont{{}}} (\bibinfo{collaboration}{CDF and D0})
  (\bibinfo{year}{2010}{\natexlab{c}}), \eprint{1007.3178}.

\bibitem[{\citenamefont{Isidori and Unterdorfer}(2004)}]{Isidori:2003ts}
\bibinfo{author}{\bibfnamefont{G.}~\bibnamefont{Isidori}} \bibnamefont{and}
  \bibinfo{author}{\bibfnamefont{R.}~\bibnamefont{Unterdorfer}},
  \bibinfo{journal}{JHEP} \textbf{\bibinfo{volume}{01}}, \bibinfo{pages}{009}
  (\bibinfo{year}{2004}), \eprint{hep-ph/0311084}.

\bibitem[{\citenamefont{D'Ambrosio et~al.}(1998)\citenamefont{D'Ambrosio,
  Isidori, and Portoles}}]{D'Ambrosio:1997jp}
\bibinfo{author}{\bibfnamefont{G.}~\bibnamefont{D'Ambrosio}},
  \bibinfo{author}{\bibfnamefont{G.}~\bibnamefont{Isidori}}, \bibnamefont{and}
  \bibinfo{author}{\bibfnamefont{J.}~\bibnamefont{Portoles}},
  \bibinfo{journal}{Phys. Lett.} \textbf{\bibinfo{volume}{B423}},
  \bibinfo{pages}{385} (\bibinfo{year}{1998}), \eprint{hep-ph/9708326}.

\bibitem[{\citenamefont{Laiho et~al.}(2010)\citenamefont{Laiho, Lunghi, and
  Van~de Water}}]{Laiho:2009eu}
\bibinfo{author}{\bibfnamefont{J.}~\bibnamefont{Laiho}},
  \bibinfo{author}{\bibfnamefont{E.}~\bibnamefont{Lunghi}}, \bibnamefont{and}
  \bibinfo{author}{\bibfnamefont{R.~S.} \bibnamefont{Van~de Water}},
  \bibinfo{journal}{Phys. Rev.} \textbf{\bibinfo{volume}{D81}},
  \bibinfo{pages}{034503} (\bibinfo{year}{2010}), \eprint{0910.2928}.

\bibitem[{\citenamefont{Valencia}(1998)}]{Valencia:1997xe}
\bibinfo{author}{\bibfnamefont{G.}~\bibnamefont{Valencia}},
  \bibinfo{journal}{Nucl.Phys.} \textbf{\bibinfo{volume}{B517}},
  \bibinfo{pages}{339} (\bibinfo{year}{1998}), \eprint{hep-ph/9711377}.

\bibitem[{\citenamefont{Ambrosino et~al.}(2009)}]{Ambrosino:2008zi}
\bibinfo{author}{\bibfnamefont{F.}~\bibnamefont{Ambrosino}}
  \bibnamefont{et~al.} (\bibinfo{collaboration}{KLOE}),
  \bibinfo{journal}{Phys.Lett.} \textbf{\bibinfo{volume}{B672}},
  \bibinfo{pages}{203} (\bibinfo{year}{2009}), \eprint{0811.1007}.

\bibitem[{\citenamefont{Ecker and Pich}(1991)}]{Ecker:1991ru}
\bibinfo{author}{\bibfnamefont{G.}~\bibnamefont{Ecker}} \bibnamefont{and}
  \bibinfo{author}{\bibfnamefont{A.}~\bibnamefont{Pich}},
  \bibinfo{journal}{Nucl.Phys.} \textbf{\bibinfo{volume}{B366}},
  \bibinfo{pages}{189} (\bibinfo{year}{1991}).

\bibitem[{\citenamefont{Serrano}(2011)}]{LHCbEPS}
\bibinfo{author}{\bibfnamefont{J.}~\bibnamefont{Serrano}}
  (\bibinfo{collaboration}{LHCb Collaboration}) (\bibinfo{year}{2011}),
  \eprint{{LHCb-TALK-2011-143}}.

\bibitem[{\citenamefont{{P. F. Harrison and H. R. Quinn
  (editors)}}(1998)}]{Harrison:1998yr}
\bibinfo{author}{\bibnamefont{{P. F. Harrison and H. R. Quinn (editors)}}}
  (\bibinfo{collaboration}{BABAR}) (\bibinfo{year}{1998}), \bibinfo{note}{{The
  BABAR physics book: Physics at an asymmetric B factory.}}

\bibitem[{\citenamefont{Aubert et~al.}(2006)}]{Aubert:2005qw}
\bibinfo{author}{\bibfnamefont{B.}~\bibnamefont{Aubert}} \bibnamefont{et~al.}
  (\bibinfo{collaboration}{BABAR}), \bibinfo{journal}{Phys. Rev. Lett.}
  \textbf{\bibinfo{volume}{96}}, \bibinfo{pages}{241802}
  (\bibinfo{year}{2006}), \eprint{hep-ex/0511015}.

\bibitem[{\citenamefont{Descotes-Genon
  et~al.}(2011)\citenamefont{Descotes-Genon, Ghosh, Matias, and
  Ramon}}]{DescotesGenon:2011yn}
\bibinfo{author}{\bibfnamefont{S.}~\bibnamefont{Descotes-Genon}},
  \bibinfo{author}{\bibfnamefont{D.}~\bibnamefont{Ghosh}},
  \bibinfo{author}{\bibfnamefont{J.}~\bibnamefont{Matias}}, \bibnamefont{and}
  \bibinfo{author}{\bibfnamefont{M.}~\bibnamefont{Ramon}}
  (\bibinfo{year}{2011}), \eprint{1104.3342}.

\bibitem[{\citenamefont{Bobeth et~al.}(2004)\citenamefont{Bobeth, Gambino,
  Gorbahn, and Haisch}}]{Bobeth:2003at}
\bibinfo{author}{\bibfnamefont{C.}~\bibnamefont{Bobeth}},
  \bibinfo{author}{\bibfnamefont{P.}~\bibnamefont{Gambino}},
  \bibinfo{author}{\bibfnamefont{M.}~\bibnamefont{Gorbahn}}, \bibnamefont{and}
  \bibinfo{author}{\bibfnamefont{U.}~\bibnamefont{Haisch}},
  \bibinfo{journal}{JHEP} \textbf{\bibinfo{volume}{0404}}, \bibinfo{pages}{071}
  (\bibinfo{year}{2004}), \eprint{hep-ph/0312090}.

\bibitem[{\citenamefont{Aubert et~al.}(2004)}]{Aubert:2004it}
\bibinfo{author}{\bibfnamefont{B.}~\bibnamefont{Aubert}} \bibnamefont{et~al.}
  (\bibinfo{collaboration}{BABAR}), \bibinfo{journal}{Phys.Rev.Lett.}
  \textbf{\bibinfo{volume}{93}}, \bibinfo{pages}{081802}
  (\bibinfo{year}{2004}), \eprint{hep-ex/0404006}.

\bibitem[{\citenamefont{Iwasaki et~al.}(2005)}]{Iwasaki:2005sy}
\bibinfo{author}{\bibfnamefont{M.}~\bibnamefont{Iwasaki}} \bibnamefont{et~al.}
  (\bibinfo{collaboration}{Belle}), \bibinfo{journal}{Phys.Rev.}
  \textbf{\bibinfo{volume}{D72}}, \bibinfo{pages}{092005}
  (\bibinfo{year}{2005}), \eprint{hep-ex/0503044}.

\bibitem[{\citenamefont{Huber et~al.}(2008)\citenamefont{Huber, Hurth, and
  Lunghi}}]{Huber:2007vv}
\bibinfo{author}{\bibfnamefont{T.}~\bibnamefont{Huber}},
  \bibinfo{author}{\bibfnamefont{T.}~\bibnamefont{Hurth}}, \bibnamefont{and}
  \bibinfo{author}{\bibfnamefont{E.}~\bibnamefont{Lunghi}},
  \bibinfo{journal}{Nucl.Phys.} \textbf{\bibinfo{volume}{B802}},
  \bibinfo{pages}{40} (\bibinfo{year}{2008}), \eprint{0712.3009}.

\bibitem[{\citenamefont{Ball and Zwicky}(2005)}]{Ball:2004ye}
\bibinfo{author}{\bibfnamefont{P.}~\bibnamefont{Ball}} \bibnamefont{and}
  \bibinfo{author}{\bibfnamefont{R.}~\bibnamefont{Zwicky}},
  \bibinfo{journal}{Phys.Rev.} \textbf{\bibinfo{volume}{D71}},
  \bibinfo{pages}{014015} (\bibinfo{year}{2005}), \eprint{hep-ph/0406232}.

\bibitem[{\citenamefont{Khodjamirian et~al.}(2011)\citenamefont{Khodjamirian,
  Mannel, Offen, and Wang}}]{Khodjamirian:2011ub}
\bibinfo{author}{\bibfnamefont{A.}~\bibnamefont{Khodjamirian}},
  \bibinfo{author}{\bibfnamefont{T.}~\bibnamefont{Mannel}},
  \bibinfo{author}{\bibfnamefont{N.}~\bibnamefont{Offen}}, \bibnamefont{and}
  \bibinfo{author}{\bibfnamefont{Y.-M.} \bibnamefont{Wang}},
  \bibinfo{journal}{Phys.Rev.} \textbf{\bibinfo{volume}{D83}},
  \bibinfo{pages}{094031} (\bibinfo{year}{2011}), \eprint{1103.2655}.

\bibitem[{\citenamefont{Feldmann et~al.}(1999)\citenamefont{Feldmann, Kroll,
  and Stech}}]{Feldmann:1998sh}
\bibinfo{author}{\bibfnamefont{T.}~\bibnamefont{Feldmann}},
  \bibinfo{author}{\bibfnamefont{P.}~\bibnamefont{Kroll}}, \bibnamefont{and}
  \bibinfo{author}{\bibfnamefont{B.}~\bibnamefont{Stech}},
  \bibinfo{journal}{Phys.Lett.} \textbf{\bibinfo{volume}{B449}},
  \bibinfo{pages}{339} (\bibinfo{year}{1999}), \eprint{hep-ph/9812269}.

\bibitem[{\citenamefont{Rosner and Stone}(2010)}]{Rosner:2010ak}
\bibinfo{author}{\bibfnamefont{J.~L.} \bibnamefont{Rosner}} \bibnamefont{and}
  \bibinfo{author}{\bibfnamefont{S.}~\bibnamefont{Stone}}
  (\bibinfo{year}{2010}), \eprint{1002.1655}.

\bibitem[{\citenamefont{Kitano et~al.}(2002)\citenamefont{Kitano, Koike, and
  Okada}}]{Kitano:2002mt}
\bibinfo{author}{\bibfnamefont{R.}~\bibnamefont{Kitano}},
  \bibinfo{author}{\bibfnamefont{M.}~\bibnamefont{Koike}}, \bibnamefont{and}
  \bibinfo{author}{\bibfnamefont{Y.}~\bibnamefont{Okada}},
  \bibinfo{journal}{Phys.Rev.} \textbf{\bibinfo{volume}{D66}},
  \bibinfo{pages}{096002} (\bibinfo{year}{2002}), \eprint{hep-ph/0203110}.

\bibitem[{\citenamefont{Dohmen et~al.}(1993)}]{Dohmen:1993mp}
\bibinfo{author}{\bibfnamefont{C.}~\bibnamefont{Dohmen}} \bibnamefont{et~al.}
  (\bibinfo{collaboration}{SINDRUM II.}), \bibinfo{journal}{Phys. Lett.}
  \textbf{\bibinfo{volume}{B317}}, \bibinfo{pages}{631} (\bibinfo{year}{1993}).

\bibitem[{\citenamefont{Bertl et~al.}(2006)}]{Bertl:2006up}
\bibinfo{author}{\bibfnamefont{W.~H.} \bibnamefont{Bertl}} \bibnamefont{et~al.}
  (\bibinfo{collaboration}{SINDRUM II}), \bibinfo{journal}{Eur. Phys. J.}
  \textbf{\bibinfo{volume}{C47}}, \bibinfo{pages}{337} (\bibinfo{year}{2006}).

\bibitem[{\citenamefont{Buras et~al.}(1990)\citenamefont{Buras, Jamin, and
  Weisz}}]{Buras:1990fn}
\bibinfo{author}{\bibfnamefont{A.~J.} \bibnamefont{Buras}},
  \bibinfo{author}{\bibfnamefont{M.}~\bibnamefont{Jamin}}, \bibnamefont{and}
  \bibinfo{author}{\bibfnamefont{P.~H.} \bibnamefont{Weisz}},
  \bibinfo{journal}{Nucl.Phys.} \textbf{\bibinfo{volume}{B347}},
  \bibinfo{pages}{491} (\bibinfo{year}{1990}).

\bibitem[{\citenamefont{Inami and Lim}(1981)}]{Inami:1980fz}
\bibinfo{author}{\bibfnamefont{T.}~\bibnamefont{Inami}} \bibnamefont{and}
  \bibinfo{author}{\bibfnamefont{C.~S.} \bibnamefont{Lim}},
  \bibinfo{journal}{Prog. Theor. Phys.} \textbf{\bibinfo{volume}{65}},
  \bibinfo{pages}{297} (\bibinfo{year}{1981}).

\bibitem[{\citenamefont{Buras et~al.}(2010)\citenamefont{Buras, Guadagnoli, and
  Isidori}}]{Buras:2010pza}
\bibinfo{author}{\bibfnamefont{A.~J.} \bibnamefont{Buras}},
  \bibinfo{author}{\bibfnamefont{D.}~\bibnamefont{Guadagnoli}},
  \bibnamefont{and} \bibinfo{author}{\bibfnamefont{G.}~\bibnamefont{Isidori}},
  \bibinfo{journal}{Phys. Lett.} \textbf{\bibinfo{volume}{B688}},
  \bibinfo{pages}{309} (\bibinfo{year}{2010}), \eprint{1002.3612}.

\bibitem[{\citenamefont{Herrlich and Nierste}(1994)}]{Herrlich:1993yv}
\bibinfo{author}{\bibfnamefont{S.}~\bibnamefont{Herrlich}} \bibnamefont{and}
  \bibinfo{author}{\bibfnamefont{U.}~\bibnamefont{Nierste}},
  \bibinfo{journal}{Nucl. Phys.} \textbf{\bibinfo{volume}{B419}},
  \bibinfo{pages}{292} (\bibinfo{year}{1994}), \eprint{hep-ph/9310311}.

\bibitem[{\citenamefont{Buras}(2011)}]{Buras:2011we}
\bibinfo{author}{\bibfnamefont{A.~J.} \bibnamefont{Buras}}
  (\bibinfo{year}{2011}), \eprint{1102.5650}.

\bibitem[{\citenamefont{Brod and Gorbahn}(2010)}]{Brod:2010mj}
\bibinfo{author}{\bibfnamefont{J.}~\bibnamefont{Brod}} \bibnamefont{and}
  \bibinfo{author}{\bibfnamefont{M.}~\bibnamefont{Gorbahn}},
  \bibinfo{journal}{Phys. Rev.} \textbf{\bibinfo{volume}{D82}},
  \bibinfo{pages}{094026} (\bibinfo{year}{2010}), \eprint{1007.0684}.

\bibitem[{\citenamefont{Dighe et~al.}(2010)\citenamefont{Dighe, Kundu, and
  Nandi}}]{Dighe:2010nj}
\bibinfo{author}{\bibfnamefont{A.}~\bibnamefont{Dighe}},
  \bibinfo{author}{\bibfnamefont{A.}~\bibnamefont{Kundu}}, \bibnamefont{and}
  \bibinfo{author}{\bibfnamefont{S.}~\bibnamefont{Nandi}},
  \bibinfo{journal}{Phys.Rev.} \textbf{\bibinfo{volume}{D82}},
  \bibinfo{pages}{031502} (\bibinfo{year}{2010}), \eprint{1005.4051}.

\bibitem[{\citenamefont{Buras}(1998)}]{Buras:1998raa}
\bibinfo{author}{\bibfnamefont{A.~J.} \bibnamefont{Buras}}, pp.
  \bibinfo{pages}{281--539} (\bibinfo{year}{1998}), \bibinfo{note}{to appear in
  'Probing the Standard Model of Particle Interactions', F.David and R. Gupta,
  eds., 1998, Elsevier Science B.V.}, \eprint{hep-ph/9806471}.

\bibitem[{\citenamefont{Asner et~al.}(2010)}]{Asner:2010qj}
\bibinfo{author}{\bibfnamefont{D.}~\bibnamefont{Asner}} \bibnamefont{et~al.}
  (\bibinfo{collaboration}{Heavy Flavor Averaging Group})
  (\bibinfo{year}{2010}), \eprint{1010.1589}.

\bibitem[{\citenamefont{Bauer and Dunn}(2011)}]{Bauer:2010dga}
\bibinfo{author}{\bibfnamefont{C.~W.} \bibnamefont{Bauer}} \bibnamefont{and}
  \bibinfo{author}{\bibfnamefont{N.~D.} \bibnamefont{Dunn}},
  \bibinfo{journal}{Phys.Lett.} \textbf{\bibinfo{volume}{B696}},
  \bibinfo{pages}{362} (\bibinfo{year}{2011}), \eprint{1006.1629}.

\bibitem[{\citenamefont{Urban et~al.}(1998)\citenamefont{Urban, Krauss,
  Jentschura, and Soff}}]{Urban:1997gw}
\bibinfo{author}{\bibfnamefont{J.}~\bibnamefont{Urban}},
  \bibinfo{author}{\bibfnamefont{F.}~\bibnamefont{Krauss}},
  \bibinfo{author}{\bibfnamefont{U.}~\bibnamefont{Jentschura}},
  \bibnamefont{and} \bibinfo{author}{\bibfnamefont{G.}~\bibnamefont{Soff}},
  \bibinfo{journal}{Nucl. Phys.} \textbf{\bibinfo{volume}{B523}},
  \bibinfo{pages}{40} (\bibinfo{year}{1998}), \eprint{hep-ph/9710245}.

\bibitem[{\citenamefont{Jegerlehner and Nyffeler}(2009)}]{Jegerlehner:2009ry}
\bibinfo{author}{\bibfnamefont{F.}~\bibnamefont{Jegerlehner}} \bibnamefont{and}
  \bibinfo{author}{\bibfnamefont{A.}~\bibnamefont{Nyffeler}},
  \bibinfo{journal}{Phys. Rept.} \textbf{\bibinfo{volume}{477}},
  \bibinfo{pages}{1} (\bibinfo{year}{2009}), \eprint{0902.3360}.

\bibitem[{\citenamefont{Chakraverty et~al.}(2001)\citenamefont{Chakraverty,
  Choudhury, and Datta}}]{Chakraverty:2001yg}
\bibinfo{author}{\bibfnamefont{D.}~\bibnamefont{Chakraverty}},
  \bibinfo{author}{\bibfnamefont{D.}~\bibnamefont{Choudhury}},
  \bibnamefont{and} \bibinfo{author}{\bibfnamefont{A.}~\bibnamefont{Datta}},
  \bibinfo{journal}{Phys. Lett.} \textbf{\bibinfo{volume}{B506}},
  \bibinfo{pages}{103} (\bibinfo{year}{2001}), \eprint{hep-ph/0102180}.

\bibitem[{\citenamefont{Adam et~al.}(2011)}]{Adam:2011ch}
\bibinfo{author}{\bibfnamefont{J.}~\bibnamefont{Adam}} \bibnamefont{et~al.}
  (\bibinfo{collaboration}{MEG collaboration}) (\bibinfo{year}{2011}),
  \eprint{1107.5547}.

\bibitem[{\citenamefont{Oakes et~al.}(2000)\citenamefont{Oakes, Yang, and
  Young}}]{Oakes:1999zi}
\bibinfo{author}{\bibfnamefont{R.~J.} \bibnamefont{Oakes}},
  \bibinfo{author}{\bibfnamefont{J.~M.} \bibnamefont{Yang}}, \bibnamefont{and}
  \bibinfo{author}{\bibfnamefont{B.-L.} \bibnamefont{Young}},
  \bibinfo{journal}{Phys. Rev.} \textbf{\bibinfo{volume}{D61}},
  \bibinfo{pages}{075007} (\bibinfo{year}{2000}), \eprint{hep-ph/9911388}.

\bibitem[{\citenamefont{Dussoni}(2009)}]{Dussoni:2009zz}
\bibinfo{author}{\bibfnamefont{S.}~\bibnamefont{Dussoni}}
  (\bibinfo{collaboration}{MEG}), \bibinfo{journal}{Nucl.Phys.Proc.Suppl.}
  \textbf{\bibinfo{volume}{187}}, \bibinfo{pages}{109} (\bibinfo{year}{2009}).

\bibitem[{\citenamefont{Adam et~al.}(2010)}]{Adam:2009ci}
\bibinfo{author}{\bibfnamefont{J.}~\bibnamefont{Adam}} \bibnamefont{et~al.}
  (\bibinfo{collaboration}{MEG collaboration}), \bibinfo{journal}{Nucl.Phys.}
  \textbf{\bibinfo{volume}{B834}}, \bibinfo{pages}{1} (\bibinfo{year}{2010}),
  \eprint{0908.2594}.

\bibitem[{\citenamefont{O'Leary et~al.}(2010)}]{O'Leary:2010af}
\bibinfo{author}{\bibfnamefont{B.}~\bibnamefont{O'Leary}} \bibnamefont{et~al.}
  (\bibinfo{collaboration}{SuperB}) (\bibinfo{year}{2010}), \eprint{1008.1541}.

\bibitem[{\citenamefont{Aushev et~al.}(2010)\citenamefont{Aushev, Bartel,
  Bondar, Brodzicka, Browder et~al.}}]{Aushev:2010bq}
\bibinfo{author}{\bibfnamefont{T.}~\bibnamefont{Aushev}},
  \bibinfo{author}{\bibfnamefont{W.}~\bibnamefont{Bartel}},
  \bibinfo{author}{\bibfnamefont{A.}~\bibnamefont{Bondar}},
  \bibinfo{author}{\bibfnamefont{J.}~\bibnamefont{Brodzicka}},
  \bibinfo{author}{\bibfnamefont{T.}~\bibnamefont{Browder}},
  \bibnamefont{et~al.} (\bibinfo{year}{2010}), \eprint{1002.5012}.

\bibitem[{\citenamefont{Ikado et~al.}(2006)}]{Ikado:2006un}
\bibinfo{author}{\bibfnamefont{K.}~\bibnamefont{Ikado}} \bibnamefont{et~al.}
  (\bibinfo{collaboration}{Belle Collaboration}),
  \bibinfo{journal}{Phys.Rev.Lett.} \textbf{\bibinfo{volume}{97}},
  \bibinfo{pages}{251802} (\bibinfo{year}{2006}), \eprint{hep-ex/0604018}.

\bibitem[{\citenamefont{Aubert et~al.}(2010)}]{:2008gx}
\bibinfo{author}{\bibfnamefont{B.}~\bibnamefont{Aubert}} \bibnamefont{et~al.}
  (\bibinfo{collaboration}{BABAR Collaboration}), \bibinfo{journal}{Phys.Rev.}
  \textbf{\bibinfo{volume}{D81}}, \bibinfo{pages}{051101}
  (\bibinfo{year}{2010}), \eprint{0809.4027}.

\bibitem[{\citenamefont{Hara et~al.}(2010)}]{Hara:2010dk}
\bibinfo{author}{\bibfnamefont{K.}~\bibnamefont{Hara}} \bibnamefont{et~al.}
  (\bibinfo{collaboration}{Belle collaboration}), \bibinfo{journal}{Phys.Rev.}
  \textbf{\bibinfo{volume}{D82}}, \bibinfo{pages}{071101}
  (\bibinfo{year}{2010}), \eprint{1006.4201}.

\bibitem[{\citenamefont{del Amo~Sanchez et~al.}(2010)}]{:2010rt}
\bibinfo{author}{\bibfnamefont{P.}~\bibnamefont{del Amo~Sanchez}}
  \bibnamefont{et~al.} (\bibinfo{collaboration}{BABAR Collaboration})
  (\bibinfo{year}{2010}), \eprint{1008.0104}.

\bibitem[{\citenamefont{Bobeth and Haisch}(2011)}]{Bobeth:2011st}
\bibinfo{author}{\bibfnamefont{C.}~\bibnamefont{Bobeth}} \bibnamefont{and}
  \bibinfo{author}{\bibfnamefont{U.}~\bibnamefont{Haisch}}
  (\bibinfo{year}{2011}), \eprint{1109.1826}.

\bibitem[{\citenamefont{Georgi and Glashow}(1974)}]{Georgi:1974sy}
\bibinfo{author}{\bibfnamefont{H.}~\bibnamefont{Georgi}} \bibnamefont{and}
  \bibinfo{author}{\bibfnamefont{S.~L.} \bibnamefont{Glashow}},
  \bibinfo{journal}{Phys. Rev. Lett.} \textbf{\bibinfo{volume}{32}},
  \bibinfo{pages}{438} (\bibinfo{year}{1974}).

\bibitem[{\citenamefont{Slansky}(1981)}]{Slansky:1981yr}
\bibinfo{author}{\bibfnamefont{R.}~\bibnamefont{Slansky}},
  \bibinfo{journal}{Phys.Rept.} \textbf{\bibinfo{volume}{79}},
  \bibinfo{pages}{1} (\bibinfo{year}{1981}).

\bibitem[{\citenamefont{Georgi and Jarlskog}(1979)}]{Georgi:1979df}
\bibinfo{author}{\bibfnamefont{H.}~\bibnamefont{Georgi}} \bibnamefont{and}
  \bibinfo{author}{\bibfnamefont{C.}~\bibnamefont{Jarlskog}},
  \bibinfo{journal}{Phys.Lett.} \textbf{\bibinfo{volume}{B86}},
  \bibinfo{pages}{297} (\bibinfo{year}{1979}).

\bibitem[{\citenamefont{Babu and Ma}(1984)}]{Babu:1984vx}
\bibinfo{author}{\bibfnamefont{K.}~\bibnamefont{Babu}} \bibnamefont{and}
  \bibinfo{author}{\bibfnamefont{E.}~\bibnamefont{Ma}},
  \bibinfo{journal}{Phys.Lett.} \textbf{\bibinfo{volume}{B144}},
  \bibinfo{pages}{381} (\bibinfo{year}{1984}).

\bibitem[{\citenamefont{Giveon et~al.}(1991)\citenamefont{Giveon, Hall, and
  Sarid}}]{Giveon:1991zm}
\bibinfo{author}{\bibfnamefont{A.}~\bibnamefont{Giveon}},
  \bibinfo{author}{\bibfnamefont{L.~J.} \bibnamefont{Hall}}, \bibnamefont{and}
  \bibinfo{author}{\bibfnamefont{U.}~\bibnamefont{Sarid}},
  \bibinfo{journal}{Phys.Lett.} \textbf{\bibinfo{volume}{B271}},
  \bibinfo{pages}{138} (\bibinfo{year}{1991}).

\bibitem[{\citenamefont{Dorsner and Fileviez~Perez}(2006)}]{Dorsner:2006dj}
\bibinfo{author}{\bibfnamefont{I.}~\bibnamefont{Dorsner}} \bibnamefont{and}
  \bibinfo{author}{\bibfnamefont{P.}~\bibnamefont{Fileviez~Perez}},
  \bibinfo{journal}{Phys.Lett.} \textbf{\bibinfo{volume}{B642}},
  \bibinfo{pages}{248} (\bibinfo{year}{2006}), \eprint{hep-ph/0606062}.

\bibitem[{\citenamefont{Dorsner and Mocioiu}(2008)}]{Dorsner:2007fy}
\bibinfo{author}{\bibfnamefont{I.}~\bibnamefont{Dorsner}} \bibnamefont{and}
  \bibinfo{author}{\bibfnamefont{I.}~\bibnamefont{Mocioiu}},
  \bibinfo{journal}{Nucl. Phys.} \textbf{\bibinfo{volume}{B796}},
  \bibinfo{pages}{123} (\bibinfo{year}{2008}), \eprint{0708.3332}.

\bibitem[{\citenamefont{Fileviez~Perez}(2007)}]{Perez:2007rm}
\bibinfo{author}{\bibfnamefont{P.}~\bibnamefont{Fileviez~Perez}},
  \bibinfo{journal}{Phys.Lett.} \textbf{\bibinfo{volume}{B654}},
  \bibinfo{pages}{189} (\bibinfo{year}{2007}), \eprint{hep-ph/0702287}.

\bibitem[{\citenamefont{Dorsner et~al.}(2009)\citenamefont{Dorsner, Fajfer,
  Kamenik, and Kosnik}}]{Dorsner:2009cu}
\bibinfo{author}{\bibfnamefont{I.}~\bibnamefont{Dorsner}},
  \bibinfo{author}{\bibfnamefont{S.}~\bibnamefont{Fajfer}},
  \bibinfo{author}{\bibfnamefont{J.~F.} \bibnamefont{Kamenik}},
  \bibnamefont{and} \bibinfo{author}{\bibfnamefont{N.}~\bibnamefont{Kosnik}},
  \bibinfo{journal}{Phys. Lett.} \textbf{\bibinfo{volume}{B682}},
  \bibinfo{pages}{67} (\bibinfo{year}{2009}), \eprint{0906.5585}.

\bibitem[{\citenamefont{Dorsner et~al.}(2006)\citenamefont{Dorsner,
  Fileviez~Perez, and Gonzalez~Felipe}}]{Dorsner:2005ii}
\bibinfo{author}{\bibfnamefont{I.}~\bibnamefont{Dorsner}},
  \bibinfo{author}{\bibfnamefont{P.}~\bibnamefont{Fileviez~Perez}},
  \bibnamefont{and}
  \bibinfo{author}{\bibfnamefont{R.}~\bibnamefont{Gonzalez~Felipe}},
  \bibinfo{journal}{Nucl. Phys.} \textbf{\bibinfo{volume}{B747}},
  \bibinfo{pages}{312} (\bibinfo{year}{2006}), \eprint{hep-ph/0512068}.

\bibitem[{\citenamefont{Buras et~al.}(1978)\citenamefont{Buras, Ellis,
  Gaillard, and Nanopoulos}}]{Buras:1977yy}
\bibinfo{author}{\bibfnamefont{A.~J.} \bibnamefont{Buras}},
  \bibinfo{author}{\bibfnamefont{J.~R.} \bibnamefont{Ellis}},
  \bibinfo{author}{\bibfnamefont{M.~K.} \bibnamefont{Gaillard}},
  \bibnamefont{and} \bibinfo{author}{\bibfnamefont{D.~V.}
  \bibnamefont{Nanopoulos}}, \bibinfo{journal}{Nucl. Phys.}
  \textbf{\bibinfo{volume}{B135}}, \bibinfo{pages}{66} (\bibinfo{year}{1978}).

\bibitem[{\citenamefont{De~Rujula et~al.}(1980)\citenamefont{De~Rujula, Georgi,
  and Glashow}}]{DeRujula:1980qc}
\bibinfo{author}{\bibfnamefont{A.}~\bibnamefont{De~Rujula}},
  \bibinfo{author}{\bibfnamefont{H.}~\bibnamefont{Georgi}}, \bibnamefont{and}
  \bibinfo{author}{\bibfnamefont{S.~L.} \bibnamefont{Glashow}},
  \bibinfo{journal}{Phys. Rev. Lett.} \textbf{\bibinfo{volume}{45}},
  \bibinfo{pages}{413} (\bibinfo{year}{1980}).

\bibitem[{\citenamefont{Fileviez~Perez
  et~al.}(2008)\citenamefont{Fileviez~Perez, Iminniyaz, and
  Rodrigo}}]{Perez:2008ry}
\bibinfo{author}{\bibfnamefont{P.}~\bibnamefont{Fileviez~Perez}},
  \bibinfo{author}{\bibfnamefont{H.}~\bibnamefont{Iminniyaz}},
  \bibnamefont{and} \bibinfo{author}{\bibfnamefont{G.}~\bibnamefont{Rodrigo}},
  \bibinfo{journal}{Phys.Rev.} \textbf{\bibinfo{volume}{D78}},
  \bibinfo{pages}{015013} (\bibinfo{year}{2008}), \eprint{0803.4156}.

\bibitem[{\citenamefont{Bajc et~al.}(2006)\citenamefont{Bajc, Melfo,
  Senjanovic, and Vissani}}]{Bajc:2005zf}
\bibinfo{author}{\bibfnamefont{B.}~\bibnamefont{Bajc}},
  \bibinfo{author}{\bibfnamefont{A.}~\bibnamefont{Melfo}},
  \bibinfo{author}{\bibfnamefont{G.}~\bibnamefont{Senjanovic}},
  \bibnamefont{and} \bibinfo{author}{\bibfnamefont{F.}~\bibnamefont{Vissani}},
  \bibinfo{journal}{Phys.Rev.} \textbf{\bibinfo{volume}{D73}},
  \bibinfo{pages}{055001} (\bibinfo{year}{2006}), \eprint{hep-ph/0510139}.

\bibitem[{\citenamefont{Minkowski}(1977)}]{Minkowski:1977sc}
\bibinfo{author}{\bibfnamefont{P.}~\bibnamefont{Minkowski}},
  \bibinfo{journal}{Phys. Lett.} \textbf{\bibinfo{volume}{B67}},
  \bibinfo{pages}{421} (\bibinfo{year}{1977}).

\bibitem[{\citenamefont{Yanagida}(1979)}]{Yanagida:1979as}
\bibinfo{author}{\bibfnamefont{T.}~\bibnamefont{Yanagida}}
  (\bibinfo{year}{1979}), \bibinfo{note}{in Proceedings of the Workshop on the
  Baryon Number of the Universe and Unified Theories, Tsukuba, Japan, 13-14 Feb
  1979}.

\bibitem[{\citenamefont{Gell-Mann et~al.}(1980)\citenamefont{Gell-Mann, Ramond,
  and Slansky}}]{GellMann:1980vs}
\bibinfo{author}{\bibfnamefont{M.}~\bibnamefont{Gell-Mann}},
  \bibinfo{author}{\bibfnamefont{P.}~\bibnamefont{Ramond}}, \bibnamefont{and}
  \bibinfo{author}{\bibfnamefont{R.}~\bibnamefont{Slansky}}
  (\bibinfo{year}{1980}), \bibinfo{note}{print-80-0576 (CERN)}.

\bibitem[{\citenamefont{Glashow}(1980)}]{Glashow:1979nm}
\bibinfo{author}{\bibfnamefont{S.~L.} \bibnamefont{Glashow}},
  \bibinfo{journal}{NATO Adv. Study Inst. Ser. B Phys.}
  \textbf{\bibinfo{volume}{59}}, \bibinfo{pages}{687} (\bibinfo{year}{1980}).

\bibitem[{\citenamefont{Mohapatra and Senjanovic}(1980)}]{Mohapatra:1979ia}
\bibinfo{author}{\bibfnamefont{R.~N.} \bibnamefont{Mohapatra}}
  \bibnamefont{and}
  \bibinfo{author}{\bibfnamefont{G.}~\bibnamefont{Senjanovic}},
  \bibinfo{journal}{Phys. Rev. Lett.} \textbf{\bibinfo{volume}{44}},
  \bibinfo{pages}{912} (\bibinfo{year}{1980}).

\bibitem[{\citenamefont{Foot et~al.}(1989)\citenamefont{Foot, Lew, He, and
  Joshi}}]{Foot:1988aq}
\bibinfo{author}{\bibfnamefont{R.}~\bibnamefont{Foot}},
  \bibinfo{author}{\bibfnamefont{H.}~\bibnamefont{Lew}},
  \bibinfo{author}{\bibfnamefont{X.~G.} \bibnamefont{He}}, \bibnamefont{and}
  \bibinfo{author}{\bibfnamefont{G.~C.} \bibnamefont{Joshi}},
  \bibinfo{journal}{Z. Phys.} \textbf{\bibinfo{volume}{C44}},
  \bibinfo{pages}{441} (\bibinfo{year}{1989}).

\bibitem[{\citenamefont{Ma}(1998)}]{Ma:1998dn}
\bibinfo{author}{\bibfnamefont{E.}~\bibnamefont{Ma}}, \bibinfo{journal}{Phys.
  Rev. Lett.} \textbf{\bibinfo{volume}{81}}, \bibinfo{pages}{1171}
  (\bibinfo{year}{1998}), \eprint{hep-ph/9805219}.

\bibitem[{\citenamefont{Bajc and Senjanovic}(2007)}]{Bajc:2006ia}
\bibinfo{author}{\bibfnamefont{B.}~\bibnamefont{Bajc}} \bibnamefont{and}
  \bibinfo{author}{\bibfnamefont{G.}~\bibnamefont{Senjanovic}},
  \bibinfo{journal}{JHEP} \textbf{\bibinfo{volume}{08}}, \bibinfo{pages}{014}
  (\bibinfo{year}{2007}), \eprint{hep-ph/0612029}.

\bibitem[{\citenamefont{Dorsner and Fileviez~Perez}(2007)}]{Dorsner:2006fx}
\bibinfo{author}{\bibfnamefont{I.}~\bibnamefont{Dorsner}} \bibnamefont{and}
  \bibinfo{author}{\bibfnamefont{P.}~\bibnamefont{Fileviez~Perez}},
  \bibinfo{journal}{JHEP} \textbf{\bibinfo{volume}{06}}, \bibinfo{pages}{029}
  (\bibinfo{year}{2007}), \eprint{hep-ph/0612216}.

\bibitem[{\citenamefont{Bajc et~al.}(2007)\citenamefont{Bajc, Nemevsek, and
  Senjanovic}}]{Bajc:2007zf}
\bibinfo{author}{\bibfnamefont{B.}~\bibnamefont{Bajc}},
  \bibinfo{author}{\bibfnamefont{M.}~\bibnamefont{Nemevsek}}, \bibnamefont{and}
  \bibinfo{author}{\bibfnamefont{G.}~\bibnamefont{Senjanovic}},
  \bibinfo{journal}{Phys. Rev.} \textbf{\bibinfo{volume}{D76}},
  \bibinfo{pages}{055011} (\bibinfo{year}{2007}), \eprint{hep-ph/0703080}.

\bibitem[{\citenamefont{Khachatryan et~al.}(2011)}]{Khachatryan:2010mp}
\bibinfo{author}{\bibfnamefont{V.}~\bibnamefont{Khachatryan}}
  \bibnamefont{et~al.} (\bibinfo{collaboration}{CMS}), \bibinfo{journal}{Phys.
  Rev. Lett.} \textbf{\bibinfo{volume}{106}}, \bibinfo{pages}{201802}
  (\bibinfo{year}{2011}), \eprint{1012.4031}.

\bibitem[{\citenamefont{Aad et~al.}(2011)}]{Aad:2011uv}
\bibinfo{author}{\bibfnamefont{G.}~\bibnamefont{Aad}} \bibnamefont{et~al.}
  (\bibinfo{collaboration}{ATLAS}) (\bibinfo{year}{2011}), \eprint{1104.4481}.

\bibitem[{\citenamefont{Lavoura et~al.}(2006)\citenamefont{Lavoura, Kuhbock,
  and Grimus}}]{Lavoura:2006dv}
\bibinfo{author}{\bibfnamefont{L.}~\bibnamefont{Lavoura}},
  \bibinfo{author}{\bibfnamefont{H.}~\bibnamefont{Kuhbock}}, \bibnamefont{and}
  \bibinfo{author}{\bibfnamefont{W.}~\bibnamefont{Grimus}},
  \bibinfo{journal}{Nucl.Phys.} \textbf{\bibinfo{volume}{B754}},
  \bibinfo{pages}{1} (\bibinfo{year}{2006}), \eprint{hep-ph/0603259}.

\end{thebibliography}

\appendix
\section{One loop contributions of $\Delta$ to $R_b$}
\label{app:Rb}
We are working in the massless limit $m_\ell=m_b=0$ and in
$d=4+\epsilon$ dimensions to regularize UV divergence. The first two
diagrams in Fig.~\ref{fig:Zbb} give the following contribution to the
1-particle irreducible~(1PI) amplitude
\begin{eqnarray}
\mc{A}^{\Delta,\mrm{1PI}} &=&\mathrm{i} g_Z \sin^2\theta_W
\frac{\sum_\ell |Y_{\ell b}|^2}{(4\pi)^2}C_{\epsilon}\Bigg[-\frac{1}{3\epsilon}+\frac{1}{6 x_Z^2}
\Big[ 9x_Z^2-2x_Z + \log(x_Z)\big(3 x_Z^2-6x_Z + 6 \log(1+x_Z)\big)\\
&&-8 f_1 +4f_2 (2x_Z-x_Z^2) +6 \mathrm{Li}_2(-x_Z) )
 - \mathrm{i} \pi  \Big( 3x_Z^2 -6x_Z+6\log(1+x_Z)\Big)\Big]
\Bigg]\mathcal A_R\,,\nonumber
\end{eqnarray}
where $C_{\epsilon}=m_\Delta^{\epsilon}/(4\pi)^{\epsilon/2} \Gamma(1-\epsilon/2)$
and $f_1$, $f_2$ are auxiliary functions defined as
\begin{eqnarray}
f_1 &=& 4\arctan\Big(\sqrt{\frac{x_Z}{4-x_Z}}\Big)\arctan\Big(\frac{\sqrt{x_Z(4-x_Z)}}{2-x_Z}\Big)
+\mathrm{Li}_2(x_Z)\\
&&+2\mathrm{Re}\Big\{
\mathrm{Li}_2\Big(\frac{x_Z}{2}+\frac{\mathrm{i}}{2}\sqrt{x_Z(4-x_Z)}\Big) -
\mathrm{Li}_2\Big(\frac{x_Z}{2} (3-x_Z)-\frac{\mathrm{i}}{2}(1-x_Z)\sqrt{x_Z(4-x_Z)}\Big)\Big\}\,,\nonumber\\
f_2&=&2\sqrt{\frac{4-x_Z}{x_Z}}\arctan\Big(\sqrt{\frac{x_Z}{4-x_Z}}\Big)\,.
\end{eqnarray}
In addition to graphs in Fig.~\ref{fig:Zbb} there are one loop
contributions of $\Delta$ to $b$-quark self-energy, corresponding to on-shell field renormalization of the
$b$-quark field 
\begin{equation}
Z_b=1+\delta_b\,, \hspace{0.5cm} \delta_b= -\frac{1}{2}\frac{\sum_\ell |Y_{\ell b}|^2}{(4\pi)^2}C_{\epsilon}\Big[-\frac{2}{\epsilon}+\frac{1}{2}\Big]\,.
\end{equation}
Combining the tree-level SM with 1PI diagrams of $\Delta$ and the field strength
renormalization we obtain the UV-finite amplitude
\begin{eqnarray}
\mc{A} &=& Z_b(\mc{A}^{\mathrm{tree}}+\mc{A}^{\Delta,\mrm{1PI}}) = \mathrm{i} g_Z\Big[\big(g_R^0 + \delta g_R\big)\mathcal A_R + g_L^0 \mathcal A_L\Big]\,,
\end{eqnarray}
where the change of right-handed coupling, $\delta g_R$, is given in
Eq.~\eqref{eq:dgr}.

\end{document}